\documentclass[sigconf]{acmart}
\usepackage{amsfonts,amssymb}

\newcommand{\ignore}[1]{}
\usepackage{xspace}

\newcommand{\ie}{\emph{i.e.,}\xspace}

\newcommand{\eg}{\emph{e.g.,}\xspace}

\newcommand{\be}{\mathbf{e}}

\newcommand{\bb}{\mathbf{b}}
\newcommand{\bs}{\mathbf{s}}

\newcommand{\bE}{\mathbf{E}}
\newcommand{\bF}{\mathbf{F}}
\newcommand{\bW}{\mathbf{W}}

\newcommand{\bM}{\mathbf{M}}

\newcommand{\bP}{\mathbf{P}}

\newcommand{\bQ}{\mathbf{Q}}

\newcommand{\bK}{\mathbf{K}}
\newcommand{\bV}{\mathbf{V}}

\usepackage{multirow}
\usepackage{multicol}
\usepackage{caption}
\usepackage{subcaption}
\usepackage{latexsym}
\usepackage{mflogo}

\AtBeginDocument{%
  \providecommand\BibTeX{{%
    \normalfont B\kern-0.5em{\scshape i\kern-0.25em b}\kern-0.8em\TeX}}}

\copyrightyear{2020}
\acmYear{2020}
\setcopyright{acmcopyright}
\acmConference[CIKM '20] {The 29th ACM International Conference on Information and Knowledge Management}{October 19--23, 2020}{Virtual Event, Ireland}
\acmBooktitle{The 29th ACM International Conference on Information and Knowledge Management (CIKM '20), October 19--23, 2020, Virtual Event, Ireland}
\acmPrice{15.00}
\acmDOI{10.1145/3340531.3411954}
\acmISBN{978-1-4503-6859-9/20/10}

\begin{document}
\fancyhead{}




\title{S$^3$-Rec: Self-Supervised Learning for Sequential Recommendation with Mutual Information Maximization}

\author{Kun Zhou$^{1\dagger}$, Hui Wang$^{1\dagger}$, Wayne Xin Zhao$^{2,3^*}$, Yutao Zhu$^{5}$, Sirui Wang$^{4}$, \and Fuzheng Zhang$^{4}$, Zhongyuan Wang$^{4}$ and Ji-Rong Wen$^{2,3}$}\thanks{$^\dagger$Equal contribution} \thanks{$^*$Corresponding author.}
\affiliation{%
 \institution{$^1$School of Information, Renmin University of China}
 \institution{$^2$Gaoling School of Artificial Intelligence, Renmin University of China}
 \institution{$^3$Beijing Key Laboratory of Big Data Management and Analysis Methods}
 \institution{$^4$Meituan-Dianping Group}
 \institution{$^5$Université de Montréal, Montréal, Québec, Canada}
}
\affiliation{%
  \institution{francis\_kun\_zhou@163.com, \{hui.wang, batmanfly, jrwen\}@ruc.edu.cn, yutao.zhu@umontreal.ca, wangsirui@meituan.com, zhfzhkris@outlook.com}
}



\begin{abstract}
Recently, significant progress has been made in sequential recommendation with deep learning. 
Existing neural sequential recommendation models usually rely on the item prediction loss to learn model parameters or data representations.
However, the model trained with this loss is prone to suffer from data sparsity problem. Since it overemphasizes the final performance, the association or fusion between context data and sequence data has not been well captured and utilized for sequential recommendation.

To tackle this problem, we propose the model \textbf{S$^3$-Rec}, which stands for \underline{S}elf-\underline{S}upervised learning for \underline{S}equential \underline{Rec}ommendation, based on the self-attentive neural architecture.  The main idea of our approach  is to utilize the intrinsic data correlation to derive self-supervision signals and enhance the data representations via pre-training methods for improving sequential recommendation.  For our task, we devise four auxiliary self-supervised objectives to learn the correlations among attribute, item, subsequence, and sequence by utilizing the mutual information maximization (MIM) principle.
MIM provides a unified way to characterize the correlation between different types of data, which is particularly suitable in our scenario. 
Extensive experiments conducted on six real-world datasets demonstrate the superiority of our proposed method over existing state-of-the-art methods, especially when only limited training data is available. Besides, we extend our self-supervised learning method to other recommendation models, which also improve their performance.
\end{abstract}




\maketitle
\section{Introduction}
Recent years have witnessed the great success of many online platforms, such as Amazon and Taobao. Within online platforms, users' behaviors are dynamic and evolving over time. Thus it is critical to capture the dynamics of sequential user behaviors for making appropriate recommendations. In order to accurately characterize user interests and provide high-quality recommendations, the task of sequential recommendation has been widely studied in the literature~\cite{DBLP:conf/icdm/Rendle10,DBLP:journals/corr/HidasiKBT15,DBLP:conf/icdm/KangM18,DBLP:conf/www/RendleFS10,DBLP:conf/wsdm/TangW18}. 

Typically, sequential recommendation methods~\cite{DBLP:journals/corr/HidasiKBT15,DBLP:conf/icdm/KangM18,DBLP:conf/www/RendleFS10,DBLP:conf/wsdm/TangW18} capture useful sequential patterns from users' historical behaviors.
Such motivation has been extensively explored with deep learning. 
Various methods using recurrent neural networks (RNNs)~\cite{DBLP:journals/corr/HidasiKBT15}, convolutional neural networks (CNNs)~\cite{DBLP:conf/wsdm/TangW18}, and self-attention mechanisms~\cite{DBLP:conf/icdm/KangM18} have been proposed to learn good representations of user preference and characterize sequential user-item interactions.
 
Furthermore, researchers have incorporated rich contextual information (such as item attributes) to neural sequential recommenders~\cite{DBLP:conf/recsys/HidasiQKT16,DBLP:conf/wsdm/HuangRZHWD19,DBLP:conf/ijcai/ZhangZLSXWLZ19}. 
It has been demonstrated that contextual information is important to consider for improving the performance of sequential recommender systems. 

Although existing methods have been shown effective to some extent,
there are two major shortcomings that are likely to affect the recommendation performance.  
First, they rely on the item prediction loss to learn the entire model. When context data is incorporated, the involved parameters are also learned through the only optimization objective. 
It has been found that such an optimization way is easy to suffer from issues such as data sparsity~\cite{DBLP:conf/cikm/SongS0DX0T19,DBLP:conf/www/RendleFS10}. Second, they overemphasize the final performance, while the association or fusion between context data and sequence data has not been well captured in data representations. As shown in increasing evidence from various fields~\cite{DBLP:conf/naacl/DevlinCLT19,DBLP:conf/iclr/HjelmFLGBTB19,DBLP:conf/iclr/KongdYLDY20}, effective data representation (\eg pre-trained contextualized embedding) has been a key factor to improve the performance of existing models or architectures. 
Therefore, there is a need to rethink the learning paradigm to develop more effective sequential recommender systems. 

To address the above issues, we borrow the idea of self-supervised learning for improving sequential recommendation. Self-supervised learning~\cite{DBLP:conf/nips/MikolovSCCD13,DBLP:conf/naacl/DevlinCLT19} is a newly emerging paradigm, which aims to let the model learn from the intrinsic structure of the raw data. A general framework of self-supervised learning is to first construct training signals directly from the raw data and then pre-train the model parameters with additionally devised optimization objectives. 
As previously discussed, limited supervision signals and ineffective data representations are the two major learning issues with existing neural sequential methods. Fortunately, self-supervised learning seems to provide a promising solution to both problems: it utilizes the intrinsic data correlation to devise auxiliary training objectives and enhances the data representations via pre-trained methods with rich self-supervised signals. However, for sequential recommendation, the context information exists in different forms or with varying intrinsics, including item, attribute, subsequence, or sequence. It is not easy to develop a unified approach to characterizing such data correlations. For this problem, we are inspired by the recently proposed mutual information maximization~(MIM) method~\cite{DBLP:journals/computer/Linsker88,DBLP:conf/iclr/HjelmFLGBTB19,DBLP:conf/iclr/KongdYLDY20,DBLP:journals/corr/abs-2007-04032}.  
It has been shown to be particularly effective to capture the correlation between different views (or parts) of the original input by maximizing the mutual information between the encoded representations of these views. 

To this end, in this paper, we propose a novel  \underline{S}elf-\underline{S}upervised learning approach to improve  \underline{S}equential \underline{Rec}ommendation with MIM, which is called \textbf{S$^3$-Rec}. 
Based on a self-attentive recommender architecture~\cite{DBLP:conf/icdm/KangM18}, we propose to first pre-train the sequential recommender with self-supervised signals and then fine-tune the model parameters according to the recommendation task. 
The major novelty lies in the pre-training stage.
In particular, we carefully devise four self-supervised optimization objectives for capturing \emph{item-attribute}, \emph{sequence-item}, \emph{sequence-attribute} and \emph{sequence-subsequence} correlations, respectively. 
These optimization objectives are developed in a unified form of MIM. 
As such, S$^3$-Rec is able to characterize the correlation in varying levels of granularity or between different forms in a general way. It is also flexible to adapt to new data types or new correlation patterns.
Via such a pre-trained method, we can effectively fuse various kinds of context data, and learn attribute-aware contextualized data representations. 
Finally, the learned data representations are fed into the neural recommender, which will be optimized according to the recommendation performance. 

To validate the effectiveness of our proposed S$^3$-Rec method, we conduct extensive experiments on six real-world recommendation datasets of different domains. Experimental results show that S$^3$-Rec achieves state-of-the-art performance compared to a number of competitive methods, especially when training data is limited.
We also show that our S$^3$-Rec is effective to adapt to other classes of neural architectures, such as GRU and CNN.

Our main contributions are summarized as follows:
(1) To the best of our knowledge, it is the first time that self-supervised learning with MIM has been applied to improve the sequential recommendation task;
(2) We propose four self-supervised optimization objectives to maximize the mutual information of context information in different forms or granularities;
(3) Extensive experiments conducted on six real-world datasets demonstrate the effectiveness of our proposed approach. 

\section{RELATED WORK}
\subsection{Sequential Recommendation}
Early works on sequential recommendation are based on the Markov Chain assumption. MC-based methods~\cite{DBLP:conf/icdm/Rendle10} estimated an item-item transition probability matrix and utilized it to predict the next item given the last interaction of a user. 
A series of works follow this line and extend it for high-order MCs~\cite{DBLP:conf/wsdm/TangW18,DBLP:conf/icdm/KangM18,DBLP:conf/recsys/HidasiQKT16}. 
With the development of the neural networks, 
Hidasi et al.~\cite{DBLP:journals/corr/HidasiKBT15} firstly introduced Gated Recurrent Units (GRU) to the session-based recommendation and a surge of following variants modified this model by introducing pair-wise loss functions~\cite{DBLP:conf/recsys/HidasiQKT16}, memory networks~\cite{DBLP:conf/wsdm/HuangRZHWD19, DBLP:conf/sigir/HuangZDWC18}, hierarchical structures~\cite{DBLP:conf/recsys/QuadranaKHC17}, copy mechanism~\cite{DBLP:conf/aaai/RenCLR0R19} and reinforcement learning~\cite{DBLP:conf/sigir/XinKAJ20}, etc.
There are also studies that leverage other architectures~\cite{DBLP:conf/wsdm/TangW18,DBLP:conf/icdm/KangM18,DBLP:conf/cikm/SunLWPLOJ19} for sequential recommendation.
However, these approaches neglect the rich attribute information about items. To tackle this problem, TransFM~\cite{DBLP:conf/recsys/PasrichaM18} utilized Factorization Machines to incorporate arbitrary real-valued features to the sequential recommendation. FDSA~\cite{DBLP:conf/ijcai/ZhangZLSXWLZ19} employed a feature-level self-attention block to leverage the attribute information about items in user history.
Despite the remarkable success of these sequential recommendation models, the correlations among attribute, item, and sequence are still not utilized and modeled sufficiently.

\subsection{Self-supervised Learning}
Self-supervised learning~\cite{DBLP:conf/nips/MikolovSCCD13,DBLP:conf/naacl/DevlinCLT19,DBLP:conf/iclr/HjelmFLGBTB19} aims at training a network on an auxiliary objective where the ground-truth samples are obtained from the raw data automatically. 
The general framework is to construct training signals directly from the correlation within the raw data and utilize them to train the model. The correlation information learned through self-supervised learning can then be easily utilized to benefit other tasks.
Several self-supervised objectives have been introduced to use non-visual but intrinsically correlated features to guide the visual feature learning~\cite{DBLP:conf/iclr/HjelmFLGBTB19}. As for language modeling~\cite{DBLP:conf/nips/MikolovSCCD13,DBLP:conf/naacl/DevlinCLT19}, it is a popular self-supervised objective for natural language processing, where the model learns to predict the next word or sentence given the previous sequences. The learned representations of words or sequences can improve the performance of downstream tasks such as machine reading comprehension~\cite{DBLP:conf/naacl/DevlinCLT19} and natural language understanding~\cite{DBLP:conf/iclr/KongdYLDY20}.

Mutual information maximization~\cite{DBLP:conf/iclr/KongdYLDY20,DBLP:journals/computer/Linsker88,DBLP:conf/iclr/HjelmFLGBTB19} is a special branch of the self-supervised learning. It is inspired by the InfoMax principle~\cite{DBLP:journals/computer/Linsker88} and has made important progress in several domains such as computer vision~\cite{DBLP:conf/iclr/HjelmFLGBTB19}, audio processing~\cite{DBLP:journals/corr/abs-1807-03748}, and nature language understanding~\cite{DBLP:conf/iclr/KongdYLDY20}. This method splits the input data into multiple (possibly overlapping) views and maximizes the mutual information between representations of these views. The views derived from other inputs are used as negative samples. 

Different from the above approaches, our work is the first to consider the correlations within the contextual information as the self-supervised signals in sequential recommendation. We maximize the mutual information among the views of the attribute, item, and sequence, which are in different levels of granularity of the contextual information. The enhanced data representations can improve recommendation performance.
\section{PRELIMINARIES}
In this section, we first formulate the sequential recommendation problem and then introduce the technique of  mutual information maximization.

\subsection{Problem Statement}
Assume that we have a set of users and items, denoted by $\mathcal{U}$ and $\mathcal{I}$, respectively, where $u\in \mathcal{U}$ denotes a user and $i\in \mathcal{I}$ denotes an item. The numbers of users and items are denoted as $|\mathcal{U}|$ and $|\mathcal{I}|$, respectively. Generally, a user $u$ has a chronologically-ordered interaction sequence with items: $\{i_1,\cdots,i_n\}$, where $n$ is the number of interactions and $i_t$ is the $t$-th item that the user $u$ has interacted with. For convenience, we use $i_{j:k}$ to denote the subsequence, \ie $i_{j:k}=\{{i_{j},\cdots,i_{k}}\}$ where $1\leq j<k\leq n$. 
Besides, each item $i$ is associated with several attributes $\mathcal{A}_i=\{a_1,\cdots,a_m\}$. For example, a song is typical with auxiliary information such as artist, album, and popularity for music recommender. All attributes constitute an attribute set $\mathcal{A}$, and the number of attributes is donated as $|\mathcal{A}|$.

Based on the above notations, we now define the task of sequential recommendation. Formally, given the historical behaviors of a user $\{i_1,\cdots,i_n\}$ and the attributes $\mathcal{A}_i$ of each item $i$, the task of sequential recommendation is to predict the next item that the user is likely to interact with at the $(n+1)$-th step.

\subsection{Mutual Information Maximization}
\label{sec:mim}
An important technique in our approach is the \emph{Mutual Information Maximization}~(MIM). It is developed on the core concept of mutual information, which 
measures dependencies between random variables.  Given two random variables $X$ and $Y$, it can be understood as how much knowing $X$ reduces the uncertainty in $Y$ or vice versa. Formally, the mutual information between $X$ and $Y$ is:
\begin{align}
    I(X,Y)=H(X)-H(X|Y)=H(Y)-H(Y|X).
\end{align}

Maximizing mutual information directly is usually intractable.
Thus we resort to a lower bound on $I(X, Y)$. One particular lower bound that has been shown to work well in practice is InfoNCE~\cite{DBLP:journals/corr/abs-1807-03748,DBLP:conf/iclr/LogeswaranL18,DBLP:conf/iclr/KongdYLDY20}, which is based on Noise Contrastive Estimation (NCE)~\cite{DBLP:journals/jmlr/GutmannH12}. InfoNCE is defined as:
\begin{align}\label{eq-InfoNCE}
    \mathbb{E}_{p(X,Y)}[f_{\theta}(x,y)-\mathbb{E}_{q(\Tilde{Y})}[\log \sum_{\Tilde{y}\in \Tilde{Y}}\exp{f_{\theta}(x,\Tilde{y})}]]+\log |\Tilde{Y}|,
\end{align}
where $x$ and $y$ are different views of an input, and $f_{\theta}$ is a function parameterized by $\theta$ 
(\eg a dot product between encoded representations of a word and its context~\cite{DBLP:conf/iclr/KongdYLDY20} or a dot product between encoded representations of an image and the local regions of the image~\cite{DBLP:conf/iclr/HjelmFLGBTB19}), and $\Tilde{Y}$ is a set of samples drawn from a proposal distribution $q(\Tilde{Y})$, which contains a positive sample $y$ and $|\Tilde{Y}|-1$ negative samples.

Note that InfoNCE is related to the cross-entropy. If $\Tilde{Y}$ always includes all possible values of the random variable $Y$ (\ie $\Tilde{Y}=Y$) and they are uniformly distributed, maximizing InfoNCE is analogous to maximize the standard cross-entropy loss:
\begin{align}
\label{infonce}
    \mathbb{E}_{p(X,Y)}[f_{\theta}(x,y)-\log \sum_{\Tilde{y}\in Y}\exp{f_{\theta}(x,\Tilde{y})}].
\end{align}
This equation shows that InfoNCE is related to maximize $p_{\theta}(y|x)$, and it approximates the summation over elements in $Y$ (\ie, the partition function) by negative sampling. 
Based on this formula, we can utilize specific $X, Y$ to maximize the mutual information between different views of the raw data,  \eg an item and its attributes, or a sequence and the items that it contains.
\section{APPROACH}

\begin{figure*}
\includegraphics[width=.9\linewidth]{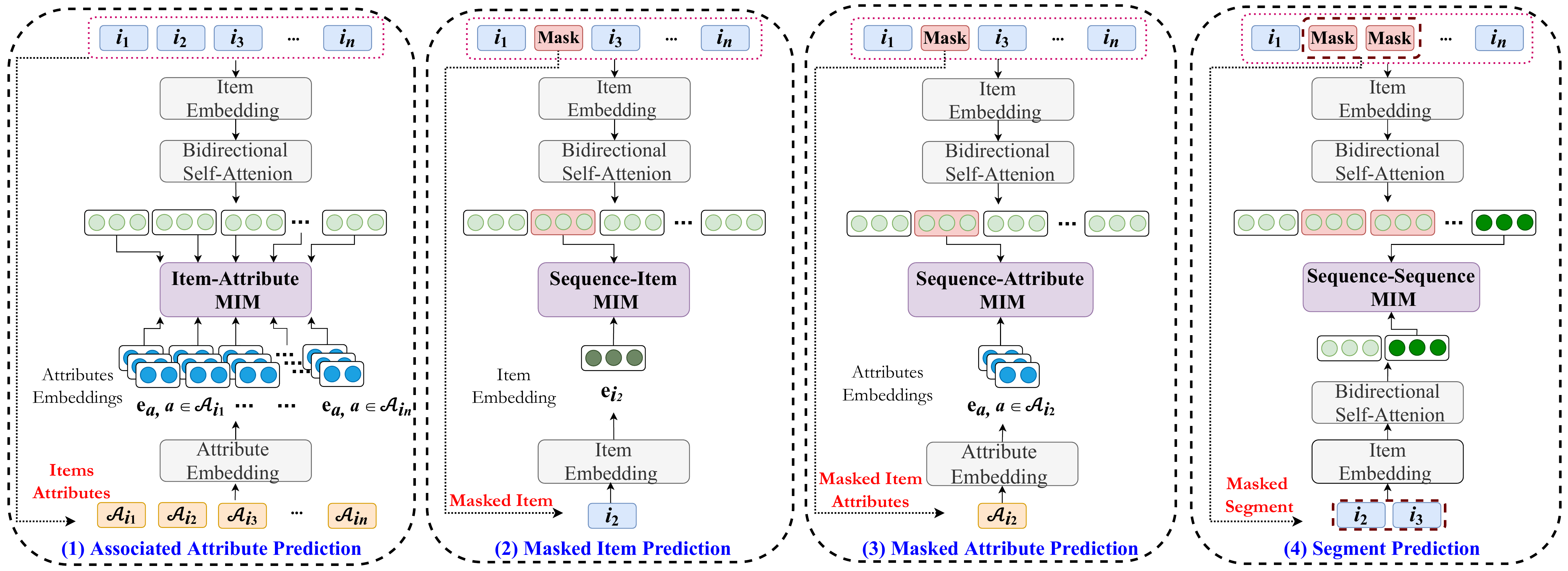}
\caption{The overview of S$^3$-Rec in the pre-training stage. We assume that the user sequence is $\{i_1, \cdots, i_n\}$ and each item $i$ is associated with several attributes $\mathcal{A}_i=\{a_1,\cdots,a_m\}$. We incorporate four self-supervised learning objectives: (1) Associated Attribute Prediction (AAP), (2) Masked Item Prediction (MIP), (3) Masked Attribute Prediction (MAP), and (4) Segment Prediction (SP). The embedding layers and bidirectional self-attention blocks are shared by the four pre-training objectives.}
\label{approach}
\end{figure*}

\subsection{Overview}
Existing studies~\cite{DBLP:conf/icdm/KangM18,DBLP:journals/corr/HidasiKBT15,DBLP:conf/recsys/HidasiQKT16,DBLP:conf/wsdm/TangW18} mainly emphasize the effect of sequential characteristics 
using an item-level optimization objective alone. Inspired by recent progress with MIM~\cite{DBLP:conf/iclr/HjelmFLGBTB19,DBLP:conf/emnlp/YehC19}, we take a different perspective to develop neural sequential recommenders by maximizing the mutual information among different views of the raw data.
 
The basic idea of our approach is to incorporate several elaborately designed self-supervised learning objectives for enhancing the original model. 
To develop such objectives, we leverage effective correlation signals reflected in the intrinsic characteristics of the input.
For our task, we consider the information in different levels of granularity, including attribute, item, segment (\ie subsequence), and sequence, which are considered as different views of the input. 
By capturing the multi-view correlation, we unify these self-supervised learning objectives with the recently proposed pre-training framework in language modeling~\cite{DBLP:conf/naacl/DevlinCLT19}.

The overview of S$^3$-Rec is presented in Fig.~\ref{approach}.
In the following sections, we first introduce the base model of our proposed approach that is developed on the Transformer architecture~\cite{DBLP:conf/icdm/KangM18}. 
Then, we will describe how we utilize the correlation signals among attributes, items, segments, and sequences to enhance the data representations based on the InfoNCE~\cite{DBLP:conf/iclr/KongdYLDY20,DBLP:journals/corr/abs-1807-03748} method. Finally, we present the discussions on our approach. 

\subsection{Base Model}
\label{base}
We develop the basic framework for sequential recommendation model by stacking the embedding layer, self-attention blocks, and the prediction layer. 

\subsubsection{Embedding Layer}
In the embedding mapping stage, we maintains an item embedding matrix $\bM_{I}\in\mathbb{R}^{|\mathcal{I}|\times d}$ and an attribute embedding matrix $\bM_{A}\in \mathbb{R}^{|\mathcal{A}|\times d}$.
The two matrices project the high-dimensional one-hot representation of an item or attribute to low-dimensional dense representations. 
Given a $n$-length item sequence, we apply a look-up operation from $\bM_{I}$ to form the input embedding matrix $\bE\in \mathbb{R}^{n\times d}$. 
Besides, we incorporate a learnable position encoding matrix $\bP\in \mathbb{R}^{n\times d}$ to enhance the input representation of the item sequence. By this means, the sequence representation $\bE_{I}\in \mathbb{R}^{n\times d}$ can be obtained by summing two embedding matrices: $\bE_{I}=\bE+\bP$. 
Since our task utilizes auxiliary context data, we also form an embedding matrix $\bE_{A}\in \mathbb{R}^{k\times d}$ for each item from the entire attribute embedding matrix $\bM_{A}$, where $k$ is the number of item attributes. 

\subsubsection{Self-Attention Block}  
Based on the embedding layer, we develop the item encoder by stacking multiple self-attention blocks. A self-attention block generally consists of two sub-layers, \ie a multi-head self-attention layer and a point-wise feed-forward network. The multi-head self-attention mechanism has been adopted for effectively extracting the information selectively from different representation subspaces. Specifically, the multi-head self-attention is defined as:
\begin{align}
    \text{MultiHeadAttn}(\bF^{l})&=[head_{1},head_{2},...,head_{h}]\bW^{O}, \\
    head_{i}&=\text{Attention}(\bF^{l}\bW_{i}^{Q}, \bF^{l}\bW_{i}^{K}, \bF^{l}\bW_{i}^{V}),
\end{align}
where the $\bF^{l}$ is the input for the $l$-th layer. When $l=0$, we set $\bF^{0}=\bE_{I}$, and the projection matrix $\bW_{i}^{Q}\in\mathbb{R}^{d\times d/h}$, $\bW_{i}^{K}\in\mathbb{R}^{d\times d/h}$, $\bW_{V}^{Q}\in \mathbb{R}^{d\times d/h}$ and $\bW^{O}\in\mathbb{R}^{d\times d}$ are the corresponding learnable parameters for each attention head. The attention function is implemented by scaled dot-product operation:
\begin{eqnarray}
    \text{Attention}(\bQ,\bK,\bV)=\text{softmax}(\frac{\bQ\bK^{\top}}{\sqrt{d/h}})\bV,
\end{eqnarray}
where $\bQ=\bF^{l}\bW^{Q}_{i}$, $\bK=\bF^{l}\bW^{K}_{i}$, and $\bV=\bF^{l}\bW^{V}_{i}$ are the linear transformations of the input embedding matrix, and $\sqrt{d/h}$ is the scale factor to avoid large values of the inner product. 

Since the multi-head attention function is mainly built on the linear projections. We endow the non-linearity of the self-attention block by applying a point-wise feed-forward network. The computation is defined as:
\begin{align}
    \bF^{l} &= [\text{FFN}(\bF^{l}_{1})^{\top};\cdots;\text{FFN}(\bF^{l}_{n})^{\top}], \label{eq:Fl}\\  
    \text{FFN}(x) &= (\text{ReLU}(x\bW_{1}+\bb_{1}))\bW_{2}+\bb_{2},
\end{align}
where $\bW_{1}$,$\bb_{1}$,$\bW_{2}$,$\bb_{2}$ are trainable parameters.

In sequential recommendation, only the information before the current time step can be utilized, thus we apply the mask operation for the output of the multi-head self-attention function to remove all connections between $\bQ_{i}$ and $\bK_{i}$. 
Inspired by BERT~\cite{DBLP:conf/naacl/DevlinCLT19}, at the pre-training stage, we remove the mask mechanism to acquire the bidirectional context-aware representation of each item in an item sequence. 
It is beneficial to incorporate context from both directions for sequence representation learning~\cite{DBLP:conf/naacl/DevlinCLT19,DBLP:conf/cikm/SunLWPLOJ19}. 

\subsubsection{Prediction Layer}
In the final layer of  S$^3$-Rec, we calculate the user's preference score for the item $i$ in the step $(t+1)$ under the context from user history as:
\begin{eqnarray}
    P(i_{t+1}=i|i_{1:t})=\be_{i}^{\top}\cdot\bF^{L}_{t},
\end{eqnarray}
where $\be_{i}$ is the representation of item $i$ from item embedding matrix $\bM_{I}$,
$\bF_{t}^{L}$ is the output of the $L$-layer self-attention block at  step $t$ and $L$ is the number of self-attention blocks.

\subsection{Self-supervised Learning with MIM}
Based on the above self-attention model, we further incorporate additional self-supervised signals with MIM to enhance the representations of input data.
We adopt a pre-training way to construct different loss functions based on the multi-view correlation. 

\subsubsection{Modeling Item-Attribute Correlation}
We first maximize the mutual information between items and attributes.
For each item, the attributes provide fine-grained information about it. Therefore, we aim to fuse item- and attribute-level information through modeling item-attribute correlation. In this way, it is expected to inject useful attribute information into item representations. 

Given an item $i$ and the attribute set $\mathcal{A}_i=\{a_{1}, \cdots, a_{k} \}$, we treat the item itself and its associated attributes as two different views.
Formally, let $\be_i$ denote the item embedding obtained by  the embedding layer, and
$\be_{a_j}$ denote the embedding for the $j$-th attribute $a_j \in \mathcal{A}_i$.
We design a loss function by the contrastive learning framework that maximizes the mutual information between the two views.
Following Eq.~\ref{infonce}, we minimize the Associated Attribute Prediction (AAP) loss  by:
\begin{align}
    L_{AAP}(i, \mathcal{A}_i)=\mathbb{E}_{a_j \in \mathcal{A}_i}[f(i, a_j)-\log \sum_{\Tilde{a}\in \mathcal{A} \setminus  \mathcal{A}_i }\exp(f(i, \Tilde{a}))],
    \label{eq:aap}
\end{align}
where we sample negative attributes $\Tilde{a}$ that enhance the association between the item $i$ and the ground-truth attributes, ``$\setminus$'' defines set subtraction operation. The function $f(\cdot,\cdot)$ is implemented with a simple bilinear network:
\begin{equation}
f(i, a_j) =  \sigma \big{(} \be_i^{\top} \cdot {\bW_{AAP}} \cdot \be_{a_j} \big{)},
\end{equation}
where $\bW_{AAP} \in \mathbb{R}^{d\times d} $ is a parameter matrix to learn and $\sigma (.)$ is the sigmoid function. 
Note that for clarity, we give the loss definition $L_{AAP}$ for a single item. It will be easy to define this loss over the entire item set. 

\subsubsection{Modeling Sequence-Item Correlation}
\label{sec:mip}
Conventional sequential recommendation models are usually trained to predict the item at the next step. This approach only considers the sequential characteristics in an item sequence from left to right. While it is noted that the entire interaction sequence is indeed observed by the model in the training process. 
Inspired by the masked language model like BERT~\cite{DBLP:conf/naacl/DevlinCLT19}, we propose to model the bidirectional information in item sequence by a Cloze task. For our task, the Cloze setting is described as below: at each training step, we randomly mask a proportion of items in the input sequence (\ie replace them with special tokens ``[\emph{mask}]''). Then we predict the masked items from the original sequence based on the surrounding context in both directions.

Therefore, the second loss we consider is to recover the actual item with the bidirectional context from the input sequences. 
For this purpose, we prepare a pre-trained version of the base model in Section~\ref{base}, which is a bidirectional Transformer architecture.
As illustration, let us mask the $t$-th item $i_t$ in a sequence $\{i_1, \cdots, i_t, \cdots, i_n\}$. We treat the rest sequence $\{i_1, \cdots, mask, \cdots, i_n\}$ as the surrounding context for $i_t$, denoted by $\mathcal{C}_{i_t}$. 
Given the surrounding context $\mathcal{C}_{i_t}$ and the masked item $i_t$, we treat them as two different views to fuse for learning data representations. Following Eq.~\ref{infonce}, we minimize the Masked Item Prediction (MIP) loss by:
\ignore{
    L_{MIP}&=\bF_{m}^{T}\bW_{M}^{I}\be(i_{t})-\log \sum_{\Tilde{i}_{m}\in \Tilde{I}}\exp(\bF_{m}^{T}\bW_{M}^{I}\be(\Tilde{i}_{t})) \\
}
\begin{align}
    L_{MIP}(\mathcal{C}_{i_t}, i_t)=f(\mathcal{C}_{i_t}, i_t)-\log [\sum_{\Tilde{i}\in \mathcal{I} \setminus \{i_t\} } f(\mathcal{C}_{i_t}, i_t) ], \label{eq:mip}
\end{align} 
where $\Tilde{i}$ denotes an irrelevant item, and $f(\cdot,\cdot)$ is implemented according to the following formula:
\begin{align}
f(\mathcal{C}_{i_t}, i_t) = \sigma \big{(} \bF_{t}^{\top} \cdot \bW_{MIP} \cdot \be_{i_{t}} \big{)},
\end{align} 
where $\bW_{MIP} \in \mathbb{R}^{d\times d}$ is a parameter matrix to learn and $\bF_{t}$ is the learned representation for the $t$-th position using the bidirectional Transformer architecture obtained in the same way as Eq.~\ref{eq:Fl}.

\subsubsection{Modeling Sequence-Attribute Correlation}
Having modeled both item-attribute and sequence-item correlations, we further consider directly fusing attribute information with sequential contexts. 
Specifically, we adopt a similar way as in Section~\ref{sec:mip} to recover the attributes of a masked item based on surrounding contexts. 
Given a masked item $i_t$, we treat its surrounding context $\mathcal{C}_{i_t}$ and its attribute set $\mathcal{A}_{i_t}$ as two different views for MIM. 
As such, we can develop the following  Masked Attribute Prediction~(MAP) loss by:

\begin{equation}
\begin{aligned}
 &L_{MAP}(\mathcal{C}_{i_t}, \mathcal{A}_{i_t}) \\= &\mathbb{E}_{a\in \mathcal{A}_{i_t}}[f( \mathcal{C}_{i_t}, a)
 -\log \sum_{\Tilde{a}  \in \mathcal{A} \setminus  \mathcal{A}_i }\exp(f(\mathcal{C}_{i_t}, \Tilde{a}))], \label{eq:map}
 \end{aligned}
\end{equation}
where $f(\cdot,\cdot)$ is implemented according to the following formula:
\begin{align}
    f(\mathcal{C}_{i_t}, a) = \sigma \big{(}  \bF_{t}^{\top} \cdot \bW_{MAP} \cdot \be_{a} \big{)},
\end{align}
where $\bW_{MAP} \in \mathbb{R}^{d \times d}$ is a parameter matrix to learn.
Note that existing methods~\cite{DBLP:conf/icdm/KangM18,DBLP:conf/wsdm/TangW18,DBLP:conf/recsys/HidasiQKT16} seldom directly model the correlation between the sequential context and attribute information. While, we would like to explicitly model the correlation to derive more meaningful supervision signals, which is useful to improve the data representations for multi-granularity information.

\subsubsection{Modeling Sequence-Segment Correlation}
As shown above, the Cloze learning strategy plays a key role in our pre-trained approach in fusing sequential contexts with target information. 
However, a major difference between item sequence with word sequence is that a single target item may not be highly related to surrounding contexts. For example, a user has bought some products just because they were on sale. 
Based on this concern, we extend the Cloze strategy from a single item to item subsequence (\ie called \emph{segment}).
Apparently, an item segment reflects more clear, stable user preference than a single item.
Therefore, we follow a similar strategy in Section~\ref{sec:mip} to recover an item subsequence from surrounding contexts. It is expected to enhance the self-supervised learning signal and improve the pre-trained performance. 

Let $i_{j_1:j_2}$ denote the subsequence from item $i_{j_1}$ to $i_{j_2}$, and 
$\mathcal{C}_{i_{j_1:j_2}}$ denote the context for $i_{j_1:j_2}$ within the entire sequence.  Similar to Eq.~\ref{eq:mip}, we can recover the missing item segment with a MIM formulation, which is so called the Segment Prediction (SP) loss as:
\begin{equation}
    \begin{aligned}
    &L_{SP}(\mathcal{C}_{i_{j_1:j_2}}, i_{j_1:j_2})\\ = &
    f(\mathcal{C}_{i_{j_1:j_2}}, i_{j_1:j_2})  - \log \sum_{\Tilde{i}_{j_1:j_2}} \exp\big( f(\mathcal{C}_{i_{j_1:j_2}}, \Tilde{i}_{j_1:j_2})  \big),  \label{eq:sp}
    \end{aligned}
\end{equation}
where $\Tilde{i}_{j_1,j_2}$ is the corrupted negative subsequence and $f(\cdot,\cdot)$ is implemented according to the following formula:
\begin{align}
f(\mathcal{C}_{i_{j_1:j_2}}, i_{j_1:j_2}) = \sigma \big{(} \bs^{\top} \cdot \bW_{SP} \cdot \tilde{\bs} \big{)},
\end{align} 
where $\bW_{SP} \in \mathbb{R}^{d\times d}$ is a parameter matrix to learn, and $\bs$ and $\tilde{\bs}$ are the learned representations for the contexts $\mathcal{C}_{i_{j_1:j_2}}$ and subsequence $i_{j_1:j_2}$, respectively.
In order to learn $\bs$ and $\tilde{\bs}$, we apply the bidirectional Transformer to obtain the state representations of the last position in a sequence. 

\subsection{Learning and Discussion}
In this part, we present the learning and related discussions of our S$^3$-Rec for sequential recommendation.

\subsubsection{Learning}
The entire procedure of S$^3$-Rec consists of two important stages, namely pre-training and fine-tuning stages. 
We adopt bidirectional and unidirectional Transformer~\cite{DBLP:conf/nips/VaswaniSPUJGKP17} architectures for the two stages, respectively.  
At the pre-trained stage, we optimize the self-supervised learning objectives by considering four different kinds of correlations (Eq.~\ref{eq:aap}, Eq.~\ref{eq:mip}, Eq.~\ref{eq:map} and Eq.~\ref{eq:sp});
at the fine-tuning stage, we utilize the learned parameters from the pre-trained stage to initialize the parameters of the unidirectional Transformer, and then utilize the left-to-right supervised signals to train the network. 
We adopt the pairwise rank loss to optimize its parameters as:
\begin{align}
\label{ft}
    L_{main} = -\sum_{u \in \mathcal{U}}\sum_{t=1}^{n}\log \sigma\bigg(P(i_{t+1}|i_{1:t})-P(i_{t+1}^{-}|i_{1:t})\bigg),
\end{align}
where we pair each ground-truth item $i_{t+1}$ with a  negative item $i_{t+1}^{-}$ that is randomly sampled. 

\subsubsection{Discussion}
Our work provides a novel self-supervised approach to capturing the intrinsic data correlation from the input as an additional signal through the pre-trained models.
This approach is quite general so that many existing methods can be included in this framework. We make a brief discussion below.   

\textbf{Feature-based approaches} such as Factorization Machine~\cite{DBLP:conf/icdm/Rendle10} and AutoInt~\cite{DBLP:conf/cikm/SongS0DX0T19} 
mainly learn data representations through the interaction of context features. 
The final prediction is made according to the actual interaction results between the user and item features. In  S$^3$-Rec, the associated attribute prediction loss $L_{AAP}$ in Eq.~\ref{eq:aap} and the masked attribute prediction   loss $L_{MAP}$ in Eq.~\ref{eq:map} have the similar effect in feature interaction. However, we do not explicitly model the interaction between attributes. Instead, we focus on capturing the association between attribute information and item/sequential contexts. 
A major difference in our work is to utilize feature interaction as additional supervision signals to enhance data representations instead of making predictions.  

\textbf{Sequential models} such as GRU4Rec~\cite{DBLP:conf/www/RendleFS10} and SASRec~\cite{DBLP:conf/icdm/KangM18} mainly focus on modeling the 
sequential dependencies between contextual items and the target item in a 
left-to-right order. S$^3$-Rec additionally incorporates a pre-trained stage that leverages four different kinds of self-supervised learning signals for enhancing data representations. In particular, the masked item prediction loss $L_{MIP}$ in Eq.~\ref{eq:mip} has a similar effect to capture sequential dependencies as in \cite{DBLP:conf/www/RendleFS10,DBLP:conf/icdm/KangM18} except that it can also utilize bidirectional sequential information. 

\textbf{Attribute-aware sequential models} such as TransFM~\cite{DBLP:conf/recsys/PasrichaM18} and FDSA~\cite{DBLP:conf/ijcai/ZhangZLSXWLZ19} leverage the contextual features to improve the sequential recommender models, in which these features are treated as auxiliary information to enhance the representation of items or sequences. 
In our S$^3$-Rec, the $L_{AAP}$ loss and $L_{MAP}$ loss aim to fuse attribute with items or sequential contexts, which is able to achieve the same effect as previous methods~\cite{DBLP:conf/recsys/PasrichaM18,DBLP:conf/ijcai/ZhangZLSXWLZ19}. Besides, the pre-trained data representations can be also applied to improve existing methods. 
\section{Experiment}

\begin{table}
    \small
	\caption{Statistics of the datasets after preprocessing.}
	\label{tab:datasets}
	\setlength{\tabcolsep}{0.6mm}{
	\begin{tabular}{lrrrrrr}
	\toprule
	Dataset & Meituan & Beauty & Sports & Toys & Yelp & LastFM \\
	\midrule
	\# Users & 13,622 & 22,363 & 25,598 & 19,412 & 30,431 & 1,090 \\
	\# Items & 20,062 & 12,101 & 18,357 & 11,924 & 20,033 & 3,646 \\
	\# Avg. Actions / User & 54.9 & 8.9 & 8.3 & 8.6 & 10.4 & 48.2 \\
	\# Avg. Actions / Item & 37.3 & 16.4 & 16.1 & 14.1 & 15.8 & 14.4 \\
	\# Actions & 747,827 & 198,502 & 296,337 & 167,597 & 316,354 & 52,551 \\
	Sparsity & 99.73\% & 99.93\% & 99.95\% & 99.93\% & 99.95\% & 98.68\% \\
	\# Attributes & 331 & 1,221 & 2,277 & 1,027 & 1,001 & 388 \\
	\# Avg. Attribute / Item & 8.8 & 5.1 & 6.0 & 4.3 & 4.8 & 31.5 \\
	\bottomrule
	\end{tabular}
	}
\end{table}

\subsection{Experimental Setup}
\subsubsection{Dataset}
We conduct experiments on six datasets collected from four real-world platforms with varying domains and sparsity levels. The statistics of these datasets after preprocessing are summarized in Table~\ref{tab:datasets}.

(1) \textbf{Meituan}\footnote{\url{https://www.meituan.com}}: this dataset consists of six-year (from \textit{Jan. 2014} to \textit{Jan. 2020}) transaction records in Beijing on the Meituan platform. We select categories, locations, and the keywords extracted from customer reviews as attributes.

(2) \textbf{Amazon Beauty, Sports, and Toys}: these three datasets are obtained from Amazon review datasets in \cite{DBLP:conf/sigir/McAuleyTSH15}. In this work, we select three subcategories: ``Beauty'', ``Sports and Outdoors'', and ``Toys and Games'', and utilize the fine-grained categories and the brands of the goods as attributes.

(3) \textbf{Yelp}\footnote{\url{https://www.yelp.com/dataset}}: this is a popular dataset for business recommendation. As it is very large, we only use the transaction records after \textit{January 1st, 2019}. We treat the categories of businesses as attributes.

(4) \textbf{LastFM}\footnote{\url{https://grouplens.org/datasets/hetrec-2011/}}: this is a music artist recommendation dataset and contains user tagging behaviors for artists. In this dataset, the tags of the artists given by the users are used as attributes.

For all datasets, we group the interaction records by users and sort them by the interaction timestamps ascendingly. Following~\cite{DBLP:conf/www/RendleFS10,DBLP:conf/ijcai/ZhangZLSXWLZ19}, we only keep the 5-core datasets, and filter unpopular items and inactive users with fewer than five interaction records.

\subsubsection{Evaluation Metrics}
We employ top-$k$ Hit Ratio (HR@$k$), top-$k$ Normalized Discounted Cumulative Gain (NDCG@$k$), and Mean Reciprocal Rank (MRR) to evaluate the performance, which are widely used in related works~\cite{DBLP:conf/www/RendleFS10,DBLP:conf/ijcai/ZhangZLSXWLZ19}.
Since HR@1 is equal to NDCG@1, we report results on HR@\{1, 5, 10\}, NGCG@\{5, 10\}, and MRR.
Following previous works~\cite{DBLP:conf/icdm/KangM18, DBLP:conf/cikm/SunLWPLOJ19, DBLP:conf/sigir/RenLLZWDW20}, we apply the \textit{leave-one-out} strategy for evaluation. Concretely, for each user interaction sequence, the last item is used as the test data, the item before the last one is used as the validation data, and the remaining data is used for training. Since the item set is large, it is time-consuming to use all items as candidates for testing. Following the common strategy~\cite{DBLP:conf/sigir/HuangZDWC18, DBLP:conf/icdm/KangM18}, we pair the ground-truth item with 99 randomly sampled negative items that the user has not interacted with. We calculate all metrics according to the ranking of the items and report the average score over all test users.

\subsubsection{Baseline Models}
We compare our proposed approach with the following eleven baseline methods:

(1) \textbf{PopRec} is a non-personalized method that ranks items according to popularity measured by the number of interactions.

(2) \textbf{FM}~\cite{DBLP:conf/icdm/Rendle10} characterizes the pairwise interactions between variables using factorized model.

(3) \textbf{AutoInt}~\cite{DBLP:conf/cikm/SongS0DX0T19} utilizes the multi-head self-attentive neural network to learn the feature interaction.

(4) \textbf{GRU4Rec}~\cite{DBLP:journals/corr/HidasiKBT15} applies GRU to model user click sequence for session-based recommendation. We represent the items using embedding vectors rather than one-hot vectors. 

(5) \textbf{Caser}~\cite{DBLP:conf/wsdm/TangW18} is a CNN-based method capturing high-order Markov Chains by applying horizontal and vertical convolutional operations for sequential recommendation. 

(6) \textbf{SASRec}~\cite{DBLP:conf/icdm/KangM18} is a self-attention based sequential recommendation model, which uses the multi-head attention mechanism to recommend the next item.

(7) \textbf{BERT4Rec}~\cite{DBLP:conf/cikm/SunLWPLOJ19} uses a Cloze objective loss for sequential recommendation by the bidirectional self-attention mechanism.

(8) \textbf{HGN}~\cite{DBLP:conf/kdd/MaKL19} is recently proposed and adopts hierarchical gating networks to capture long-term and short-term user interests.

(9) \textbf{GRU4Rec$_F$}~\cite{DBLP:conf/recsys/HidasiQKT16} is an improved version of GRU4Rec, which leverages attributes to improve the performance.

(10) \textbf{SASRec$_F$} is our extension of SASRec, which concatenates the representations of item and attribute as the input to the model.

(11) \textbf{FDSA}~\cite{DBLP:conf/ijcai/ZhangZLSXWLZ19} constructs a feature sequence and uses a feature-level self-attention block to model the feature transition patterns. This is the state-of-the-art model in sequential recommendation.

\subsubsection{Implementation Details}
For Caser and HGN, we use the source code provided by their authors. For other methods, we implement them by PyTorch. All hyper-parameters are set following the suggestions from the original papers.

For our proposed S$^{3}$-Rec, we set the number of the self-attention blocks and the attention heads as 2. The dimension of the embedding is 64, and the maximum sequence length is 50 (following~\cite{DBLP:conf/icdm/KangM18}). Note that our training phase contains two stages (\ie pre-training and fine-tuning stage), the learned parameters in the pre-training stage are used to initialize the embedding layers and self-attention layers of our model in the fine-tuning stage.

In the pre-training stage, the mask proportion of item is set as 0.2 and the weights for the four losses (\ie AAP, MIP, MAP, and SP) are set as 0.2, 1.0, 1.0, and 0.5, respectively, based on our empirical experiments. 
We use the Adam optimizer~\cite{DBLP:journals/corr/KingmaB14} with a learning rate of 0.001, where the batch size is set as 200 and 256 in the pre-training and the fine-tuning stage, respectively. We pre-train our model for 100 epochs and fine-tune it on the recommendation task. The code and data set are available at the link: \textcolor{blue}{\url{https://github.com/RUCAIBox/CIKM2020-S3Rec}}~\footnote{To further verify the effectiveness of our method, we have performed the experiments that rank the ground-truth item with all the items as candidates. The complete results are shown on our project website at this link.}.

\subsection{Experimental Results}
\begin{table*}[t!]
    \small
	\caption{Performance comparison of different methods on six datasets. The best performance and the second best performance methods are denoted in bold and underlined fonts respectively. ``$*$'' indicates the statistical significance for $p<0.01$ compared to the best baseline method.}
	\label{tab:main_table_cikm2020}
	\setlength{\tabcolsep}{1.3mm}{
	\begin{tabular}{llccccccccccclr}
	\toprule
		Datasets & Metric & PopRec &FM& AutoInt &GRU4Rec & Caser &SASRec &BERT4Rec &HGN &GRU4Rec$_F$ &SASRec$_F$ &FDSA & S$^3$-Rec & Improv. \\
	\midrule
\multirow{6} * {Meituan}
 &HR@1 &0.0946 &0.1084 &0.0804 &0.1194 &0.1368 &\underline{0.1797} &0.1381 &0.1603 &0.1436 &0.1746 &0.1778 &\textbf{0.2040$^*$} &13.52\%\\
 &HR@5 &0.2660 &0.3218 &0.2662 &0.3382 &0.3812 &0.4524 &0.3985 &0.4110 &0.3799 &0.4386 &\underline{0.4595} &\textbf{0.4925$^*$} &7.18\%\\
 &NDCG@5 &0.1813 &0.2170 &0.1739 &0.2303 &0.2619 &0.3207 &0.2713 &0.2887 &0.2639 &0.3098 &\underline{0.3236} &\textbf{0.3527$^*$} &8.99\%\\
 &HR@10 &0.3863 &0.4709 &0.4077 &0.4881 &0.5267 &0.6053 &0.5514 &0.5573 &0.5378 &0.5962 &\underline{0.6164} &\textbf{0.6368$^*$} &3.31\%\\
 &NDCG@10 &0.2200 &0.2651 &0.2194 &0.2787 &0.3090 &0.3700 &0.3208 &0.3359 &0.3149 &0.3607 &\underline{0.3743} &\textbf{0.3994$^*$} &6.71\%\\
 &MRR &0.1923 &0.2242 &0.1854 &0.2359 &0.2617 &0.3146 &0.2689 &0.2863 &0.2666 &0.3064 &\underline{0.3167} &\textbf{0.3421$^*$} &8.02\%\\
\hline
\multirow{6} * {Beauty}
 &HR@1 &0.0678 &0.0405 &0.0447 &0.1337 &0.1337 &\underline{0.1870} &0.1531 &0.1683 &0.1702 &0.1778 &0.1840 &\textbf{0.2192$^*$} &17.22\%\\
 &HR@5 &0.2105 &0.1461 &0.1705 &0.3125 &0.3032 &0.3741 &0.3640 &0.3544 &0.3727 &0.3863 &\underline{0.4010} &\textbf{0.4502$^*$} &12.27\%\\
 &NDCG@5 &0.1391 &0.0934 &0.1063 &0.2268 &0.2219 &0.2848 &0.2622 &0.2656 &0.2759 &0.2870 &\underline{0.2974} &\textbf{0.3407$^*$} &14.56\%\\
 &HR@10 &0.3386 &0.2311 &0.2872 &0.4106 &0.3942 &0.4696 &0.4739 &0.4503 &0.4753 &0.4843 &\underline{0.5096} &\textbf{0.5506$^*$} &8.05\%\\
 &NDCG@10 &0.1803 &0.1207 &0.1440 &0.2584 &0.2512 &0.3156 &0.2975 &0.2965 &0.3090 &0.3185 &\underline{0.3324} &\textbf{0.3732$^*$} &12.27\%\\
 &MRR &0.1558 &0.1096 &0.1226 &0.2308 &0.2263 &0.2852 &0.2614 &0.2669 &0.2751 &0.2844 &\underline{0.2943} &\textbf{0.3340$^*$} &13.49\%\\
\hline
\multirow{6} * {Sports}
 &HR@1 &0.0763 &0.0489 &0.0644 &0.1160 &0.1135 &0.1455 &0.1255 &0.1428 &0.1466 &0.1573 &\underline{0.1585} &\textbf{0.1841$^*$} &16.15\%\\
 &HR@5 &0.2293 &0.1603 &0.1982 &0.3055 &0.2866 &0.3466 &0.3375 &0.3349 &0.3547 &0.3730 &\underline{0.3855} &\textbf{0.4267$^*$} &10.69\%\\
 &NDCG@5 &0.1538 &0.1048 &0.1316 &0.2126 &0.2020 &0.2497 &0.2341 &0.2420 &0.2535 &0.2683 &\underline{0.2756} &\textbf{0.3104$^*$} &12.63\%\\
 &HR@10 &0.3423 &0.2491 &0.2967 &0.4299 &0.4014 &0.4622 &0.4722 &0.4551 &0.4758 &0.4912 &\underline{0.5136} &\textbf{0.5614$^*$} &9.31\%\\
 &NDCG@10 &0.1902 &0.1334 &0.1633 &0.2527 &0.2390 &0.2869 &0.2775 &0.2806 &0.2925 &0.3064 &\underline{0.3170} &\textbf{0.3538$^*$} &11.61\%\\
 &MRR &0.1660 &0.1202 &0.1435 &0.2191 &0.2100 &0.2520 &0.2378 &0.2469 &0.2549 &0.2680 &\underline{0.2748} &\textbf{0.3071$^*$} &11.75\%\\
\hline
\multirow{6} * {Toys}
 &HR@1 &0.0585 &0.0257 &0.0448 &0.0997 &0.1114 &\underline{0.1878} &0.1262 &0.1504 &0.1673 &0.1797 &0.1717 &\textbf{0.2003$^*$} &6.66\%\\
 &HR@5 &0.1977 &0.0978 &0.1471 &0.2795 &0.2614 &0.3682 &0.3344 &0.3276 &0.3695 &0.3927 &\underline{0.3994} &\textbf{0.4420$^*$} &10.67\%\\
 &NDCG@5 &0.1286 &0.0614 &0.0960 &0.1919 &0.1885 &0.2820 &0.2327 &0.2423 &0.2719 &\underline{0.2911} &0.2903 &\textbf{0.3270$^*$} &12.33\%\\
 &HR@10 &0.3008 &0.1715 &0.2369 &0.3896 &0.3540 &0.4663 &0.4493 &0.4211 &0.4782 &0.4981 &\underline{0.5129} &\textbf{0.5530$^*$} &7.82\%\\
 &NDCG@10 &0.1618 &0.0850 &0.1248 &0.2274 &0.2183 &0.3136 &0.2698 &0.2724 &0.3070 &0.3252 &\underline{0.3271} &\textbf{0.3629$^*$} &10.94\%\\
 &MRR &0.1430 &0.0819 &0.1131 &0.1973 &0.1967 &0.2842 &0.2338 &0.2454 &0.2717 &\underline{0.2886} &0.2863 &\textbf{0.3202$^*$} &10.95\%\\
\hline
\multirow{6} * {Yelp}
 &HR@1 &0.0801 &0.0624 &0.0731 &0.2053 &0.2188 &0.2375 &0.2405 &\underline{0.2428} &0.2293 &0.2301 &0.2198 &\textbf{0.2591$^*$} &6.71\%\\
 &HR@5 &0.2415 &0.2036 &0.2249 &0.5437 &0.5111 &0.5745 &\underline{0.5976} &0.5768 &0.5858 &0.5937 &0.5728 &\textbf{0.6085$^*$} &1.82\%\\
 &NDCG@5 &0.1622 &0.1333 &0.1501 &0.3784 &0.3696 &0.4113 &\underline{0.4252} &0.4162 &0.4137 &0.4178 &0.4014 &\textbf{0.4401$^*$} &3.50\%\\
 &HR@10 &0.3609 &0.3153 &0.3367 &0.7265 &0.6661 &0.7373 &0.7597 &0.7411 &0.7574 &\underline{0.7706} &0.7555 &\textbf{0.7725} &0.25\%\\
 &NDCG@10 &0.2007 &0.1692 &0.1860 &0.4375 &0.4198 &0.4642 &\underline{0.4778} &0.4695 &0.4694 &0.4751 &0.4607 &\textbf{0.4934$^*$} &3.26\%\\
 &MRR &0.1740 &0.1470 &0.1616 &0.3630 &0.3595 &0.3927 &\underline{0.4026} &0.3988 &0.3929 &0.3962 &0.3834 &\textbf{0.4190$^*$} &4.07\%\\
\hline
\multirow{6} * {LastFM}
 &HR@1 &0.0725 &0.0183 &0.0349 &0.0642 &0.0899 &0.1211 &0.1220 &0.0908 &\underline{0.1385} &0.1147 &0.0936 &\textbf{0.1743$^*$} &25.85\%\\
 &HR@5 &0.1982 &0.0954 &0.1550 &0.1817 &0.2982 &0.3385 &\underline{0.3569} &0.2872 &0.3202 &0.3073 &0.2624 &\textbf{0.4523$^*$} &26.73\%\\
 &NDCG@5 &0.1350 &0.0552 &0.0946 &0.1228 &0.1960 &0.2330 &\underline{0.2409} &0.1896 &0.2301 &0.2113 &0.1766 &\textbf{0.3156$^*$} &31.01\%\\
 &HR@10 &0.3037 &0.1578 &0.2596 &0.2817 &0.4431 &0.4706 &\underline{0.4991} &0.4193 &0.4670 &0.4569 &0.4055 &\textbf{0.5835$^*$} &16.91\%\\
 &NDCG@10 &0.1687 &0.0753 &0.1285 &0.1550 &0.2428 &0.2755 &\underline{0.2871} &0.2324 &0.2775 &0.2594 &0.2225 &\textbf{0.3583$^*$} &24.80\%\\
 &MRR &0.1506 &0.0743 &0.1122 &0.1405 &0.2033 &0.2364 &\underline{0.2424} &0.1983 &0.2410 &0.2201 &0.1884 &\textbf{0.3072$^*$} &26.73\%\\

\bottomrule
	\end{tabular}
	}
\end{table*}

\begin{figure*}[t!]
    \centering
    \includegraphics[width=\linewidth]{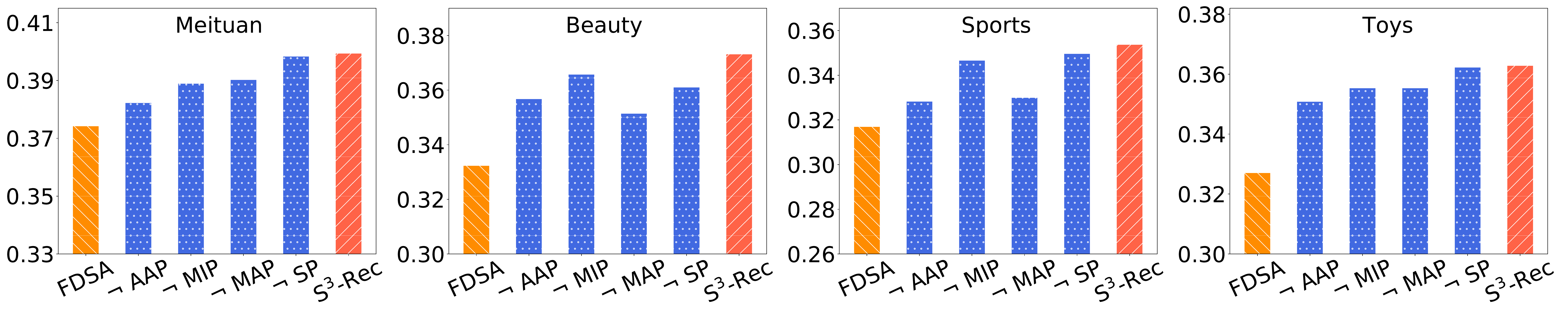}
    \caption{Ablation study of our approach on four datasets  (NDCG@10). ``$\neg$'' indicates that the corresponding objective is removed in the pre-training stage, while the rest objectives are kept.}
    \label{fig-loss-ablation}
\end{figure*}

The results of different methods on all datasets are shown in Table~\ref{tab:main_table_cikm2020}. Based on the results, we can find:

For three non-sequential recommendation baselines, the performance order is consistent across all datasets, \ie \textit{PopRec $>$ AutoInt $>$ FM}.
Due to the ``rich-gets-richer'' effect in product adoption, PopRec is a robust baseline.
AutoInt performs better than FM on most datasets because the multi-head self-attention mechanism has a stronger capacity to model attributes.
However, the performance of AutoInt is worse than that of FM on Meituan dataset. A potential reason is that the multi-head self-attention may incorporate more noise from the attributes since they are keywords extracted from the reviews on Meituan platform.
In general, non-sequential recommendation methods perform worse than sequential recommendation methods, since the sequential pattern is important to consider in our task.

\begin{figure}[t!]
    \centering
    \includegraphics[width=.85\linewidth]{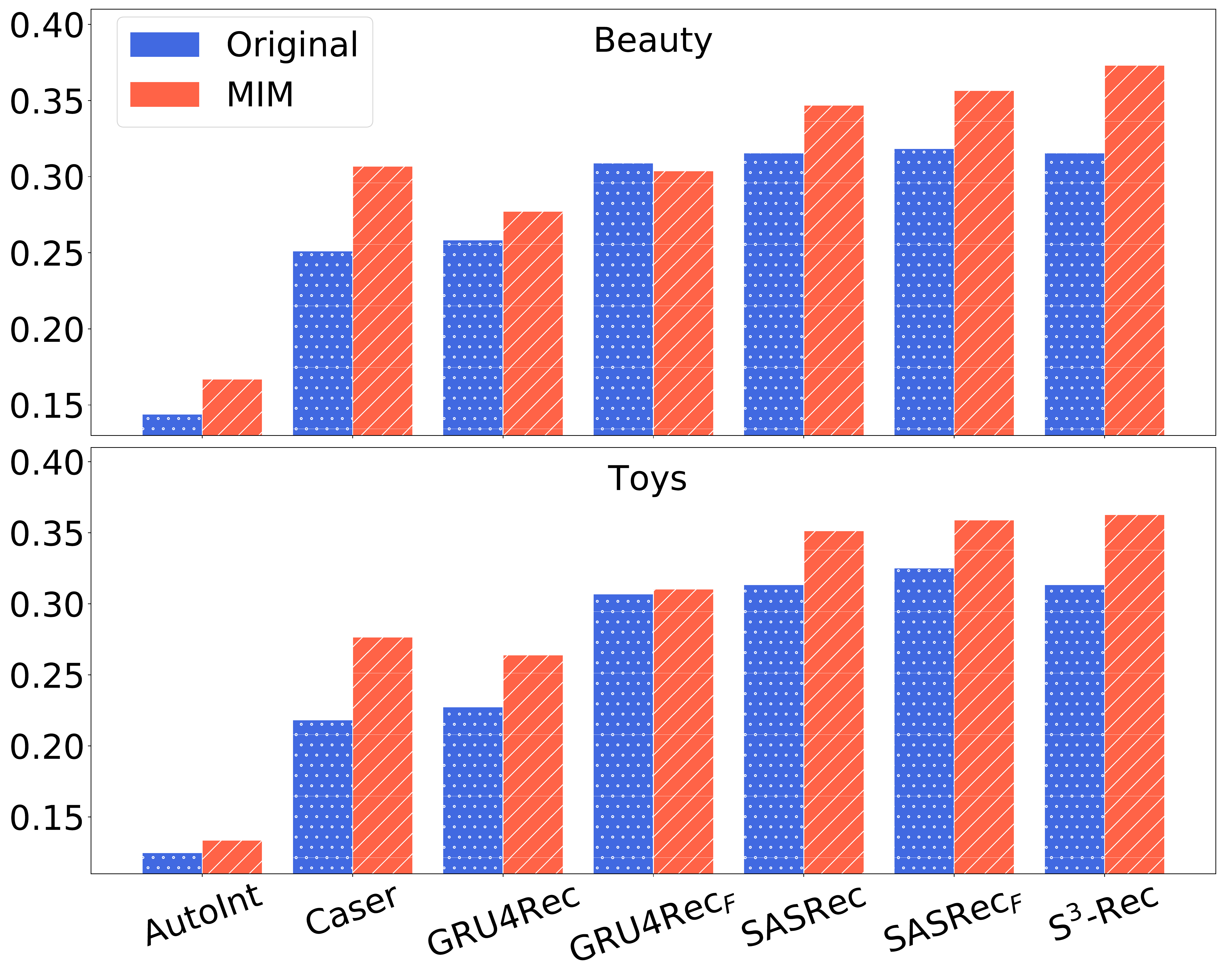}    \label{beauty-other}
    \caption{Performance (NDCG@10) comparison of different models enhanced by our self-supervised learning approach on Beauty and Toys datasets.}
\label{fig-other-extend}
\end{figure}

As for sequential recommendation baseline methods, SASRec and BERT4Rec utilize the unidirectional and bidirectional self-attention mechanism respectively, and achieve better performance than GRU4Rec and Caser. 
It indicates that self-attentive architecture is particularly suitable for modeling sequential data. However, their improvements are not stable when training with the conventional next-item prediction loss.
Besides, HGN achieves comparable performance with SASRec and BERT4Rec. This indicates the hierarchical gating network can well model the relations between closely relevant items.
However, when directly injecting the attribute information into GRU4Rec and SASRec (\ie GRU4Rec$_F$ and SASRec$_F$), the performance improvement is not consistent. This method yields improvement on Beauty, Sports, Toys, and Yelp datasets, but has a negative influence on other datasets. One possible reason is that simply concatenating item representations and its attributes representations cannot effectively fuse the two kinds of information.
In most cases, FDSA achieves the best performance among all baselines.
This suggests that the feature-level self-attention blocks can capture useful sequential feature interaction patterns.

Finally, by comparing our approach with all the baselines, it is clear to see that S$^3$-Rec performs consistently better than them by a large margin on six datasets. 
Different from these baselines, we adopt the self-supervised learning to enhance the representations of the attribute, item, and sequence for the recommendation task, which incorporates four pre-training objectives to model multiple data correlations by MIM.
This result also shows that the self-supervised approach is effective to improve the performance of the self-attention architecture for sequential recommendation.

\begin{figure}[t!]
    \centering
    \begin{subfigure}[b]{0.49\linewidth}
        \centering
        \includegraphics[width=\textwidth]{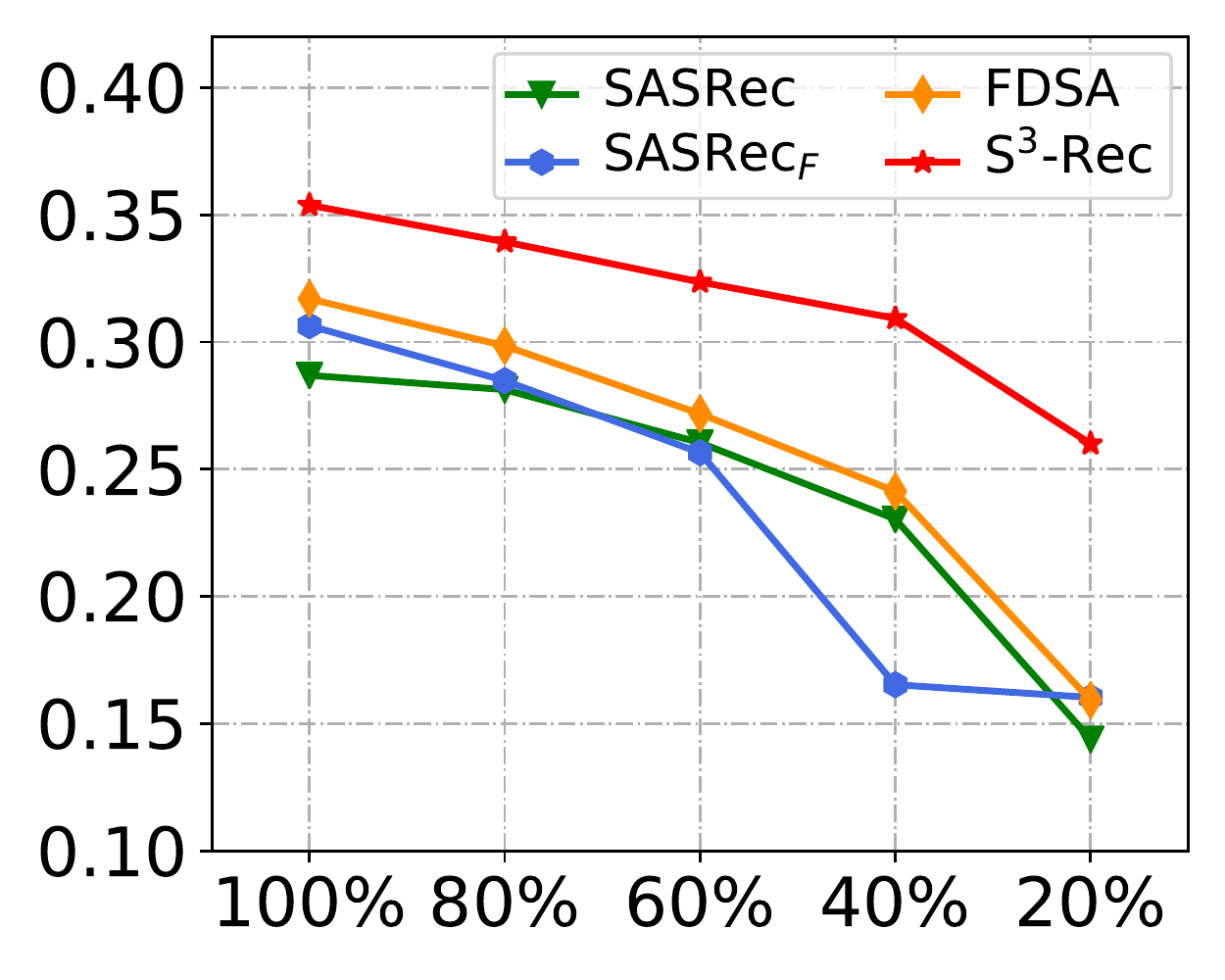}
        \caption{Sports}
        \label{sports-data-ndcg}
    \end{subfigure}
    \begin{subfigure}[b]{0.49\linewidth}
        \centering
        \includegraphics[width=\textwidth]{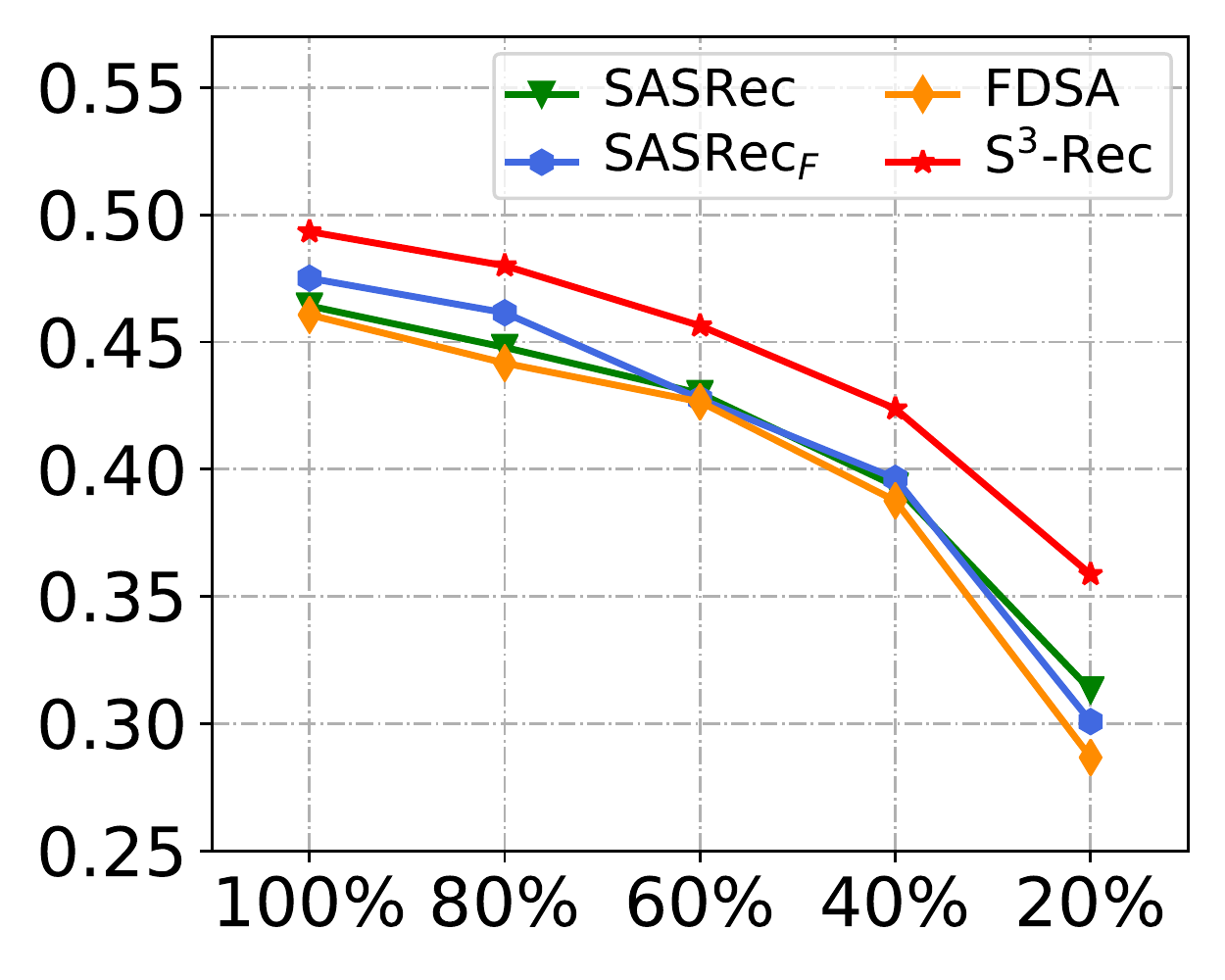}
        \caption{Yelp}
        \label{yelp-data-ndcg}
    \end{subfigure}
    \caption{Performance (NDCG@10) comparison w.r.t. different sparsity levels on Sport and Yelp datasets.}
    \vspace{-0.2cm}
\label{fig-data-amount}
\end{figure}

\subsection{Further Analysis}
Next, we continue to study whether S$^3$-Rec works well in more detailed analysis.

\subsubsection{Ablation Study}
Our proposed self-supervised approach S$^3$-Rec designs four pre-training objectives based on MIM. 
To verify the effectiveness of each objective, we conduct the ablation study on Meituan, Beauty, Sports, and Toys datasets to analyze the contribution of each objective. NDCG@10 is adopted for this evaluation. The results from the best baseline FDSA are also provided for comparison.

From the results in Fig.~\ref{fig-loss-ablation}, we can observe that removing any self-supervised objective would lead to the performance decrease. It indicates all the objectives are useful to improve the recommendation performance. Besides, the importance of these objectives is varying on different datasets. Overall, the AAP (Associated Attribute Prediction) and the MAP (Masked Attribute Prediction) are more important than the other objectives. Removing each of them yields a larger drop of performance on all datasets. One possible reason is that these two objectives enhance the representations of item and sequence with the attributes information.

It is clearly seen that all model variants are better than the best baseline FDSA, which is trained only with next-item predication loss.

\subsubsection{Applying Self-Supervised Learning to Other Models}
Since self-supervised learning itself is a learning paradigm, it can generally apply to various models. Thus, in this part, we conduct an experiment to examine whether our method can bring improvements to other models. We use the self-supervised approach to pre-training some baseline models on Beauty and Toys datasets.
For GRU4Rec, GRU4Rec$_F$, SASRec, and SASRec$_F$, we directly apply our pre-training objectives to improve them.
It is worth noting that GRU4Rec and SASRec are unidirectional models, so we maintain the unidirectional encoder layer in the pre-training stage.
For AutoInt and Caser, since their architectures do not support some of the pre-training objectives\footnote{Because their base models do not support the \textit{mask} operations.}, we only utilize the pre-trained parameters to  initialize the parameters of the embedding layers. 

The results of NDCG@10 on Beauty and Toys datasets are shown in Fig.~\ref{fig-other-extend}. First, after pre-training by our approach, all the baselines achieve better performance. This shows that self-supervised learning can also be applied to improve their performance. 
Second, S$^3$-Rec outperforms all the baselines after pre-training. This is because our model adopts the bidirectional Transformer encoder in the pre-training stage, which is more suitable for our approach. 
Third, we can see the GRU-based models achieve less improvement than the other models. One possible reason is that RNN-based architecture limits the potential of self-supervised learning.

\begin{figure}[t!]
    \centering
    \begin{subfigure}[b]{0.49\linewidth}
        \centering
        \includegraphics[width=\textwidth]{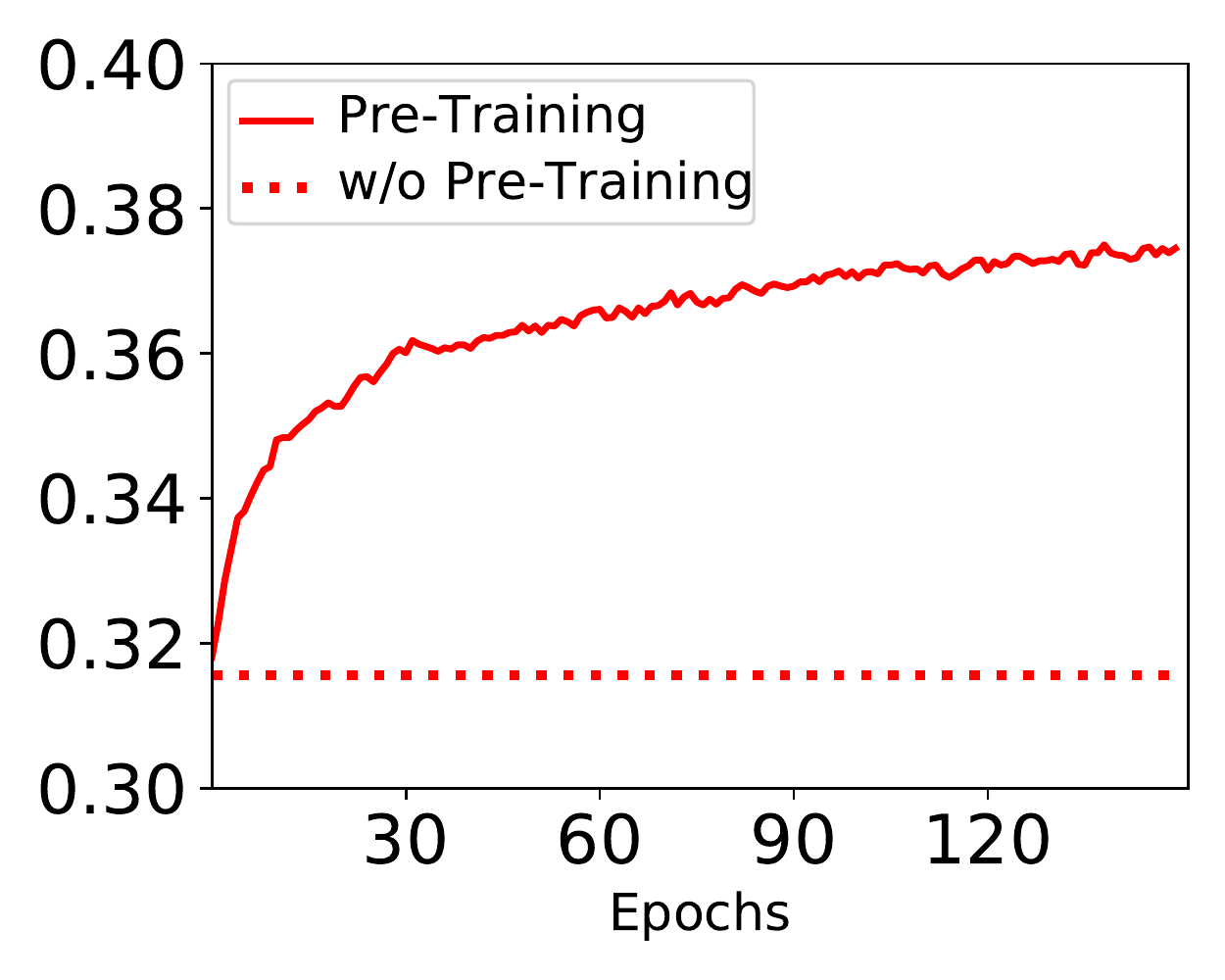}
        \caption{Beauty}
        \label{beauty-pre-epoch}
    \end{subfigure}
    \begin{subfigure}[b]{0.49\linewidth}
        \centering
        \includegraphics[width=\textwidth]{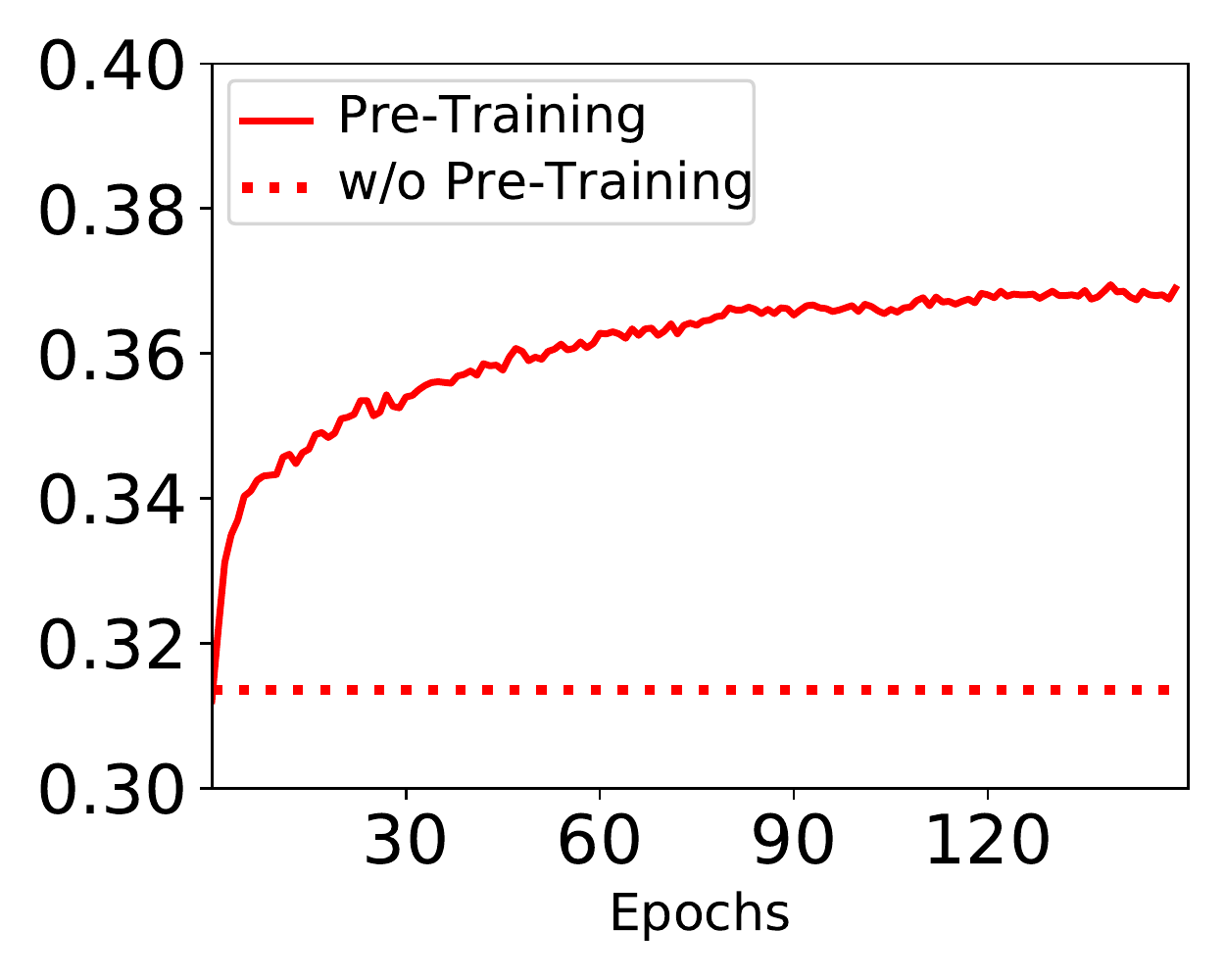}
        \caption{Toys}
        \label{toys-pre-epoch}
    \end{subfigure}
    \caption{Performance (NDCG@10) comparison w.r.t. different numbers of pre-training epochs on Beauty and Toys datasets.}
\label{fig-pre-epoch}
\end{figure}

\subsubsection{Performance Comparison w.r.t. the Amount of Training Data}
Conventional recommendation systems require a considerable amo- unt of training data, thus they are likely to suffer from the cold start problem in real-world applications. 
This problem can be alleviated by our method because the proposed self-supervised learning approach can better utilize the data correlation from input. 
We simulate the data sparsity scenarios by using different proportions of the full dataset, \ie 20\%, 40\%, 60\%, 80\%, and 100\%. 

Fig.~\ref{fig-data-amount} shows the evaluation results on Sports and Yelp datasets. As  we can see, the performance substantially drops  when less training data is used. 
While, S$^3$-Rec is consistently better than baselines in all cases, especially in an extreme sparsity level (20\%).
This observation implies that S$^3$-Rec is able to make better use of the data with the self-supervised method, which alleviates the influence of data sparsity problem for sequential recommendation to some extent.

\subsubsection{Performance Comparison w.r.t. the Number of Pre-training Epochs}
Our approach consists of a pre-training stage and a fine-tuning stage. In the pre-training stage, our model can learn the enhanced representations of the attribute, item, subsequence, and sequence for the recommendation task. The number of pre-training epochs affects the performance of the recommendation task. To investigate this, we pre-train our model with a varying number of epochs and fine-tune it on the recommendation task.

Fig.~\ref{fig-pre-epoch} presents the results on Beauty and Toys datasets.
The horizontal dash lines represent the performance without pre-training.
We can see that our model benefits mostly from the first 20 pre-training epochs. And after that, the performance improves slightly.
Based on this observation, we can conclude that the correlations among different views (\ie the attribute, item, subsequence, and sequence) can be well-captured by our self-supervised learning approach through pre-training within a small number of epochs. So that the enhanced data representations can improve the performance of sequential recommendation.

\subsubsection{Convergence Speed Comparison}
\begin{figure}[t]
    \centering
    \begin{subfigure}[b]{0.49\linewidth}
        \centering
        \includegraphics[width=\textwidth]{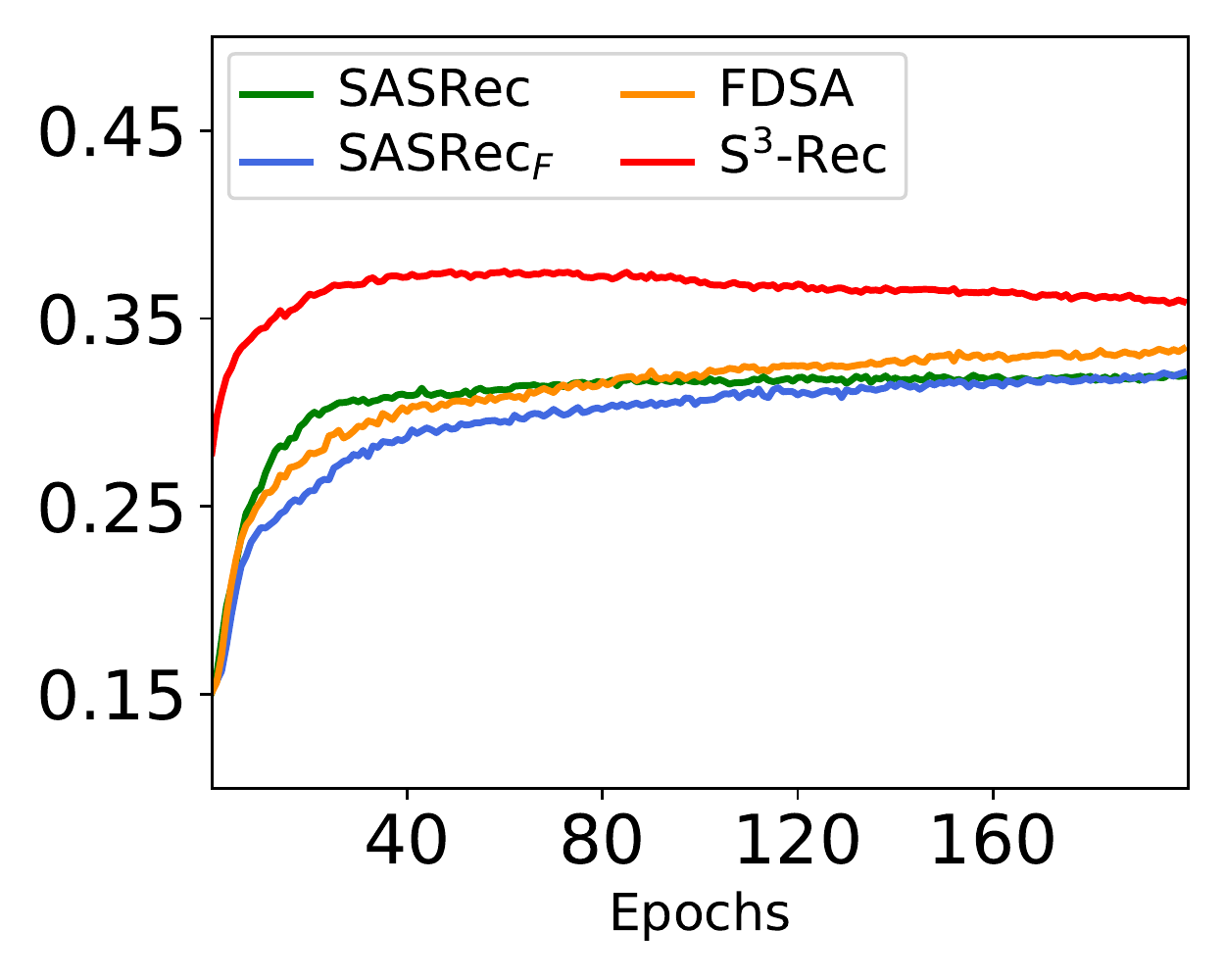}
        \caption{Beauty}
        \label{beauty-conv-curve}
    \end{subfigure}
    \begin{subfigure}[b]{0.49\linewidth}
        \centering
        \includegraphics[width=\textwidth]{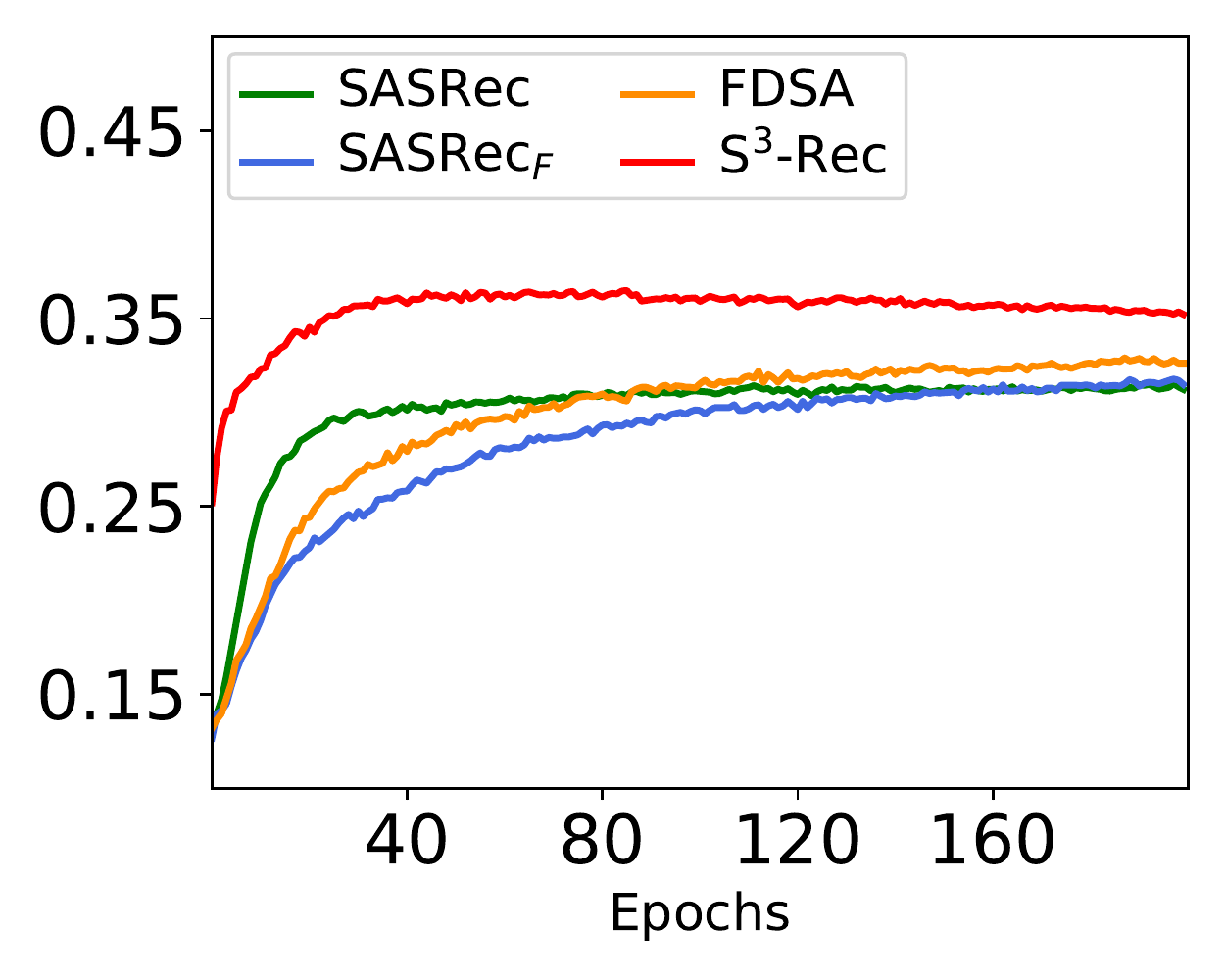}
        \caption{Toys}
        \label{toys-conv-curve}
    \end{subfigure}
    \caption{Performance tuning (NDCG@10) of our approach and other baselines with the increasing iterations in the fine-tuning stage.}
\label{fig-conv-curve}
\end{figure}
After obtaining the enhanced representations of the attribute, item, and sequence, we fine-tune our model on the recommendation task. To examine the convergence speed on the final recommendation task, we gradually increase the number of epochs for the fine-tuning stage and compare the performance of our model and  other baselines.

Fig.~\ref{fig-conv-curve} shows the results on Beauty and Toys datasets. It can be observed that our model converges quickly and achieves the best performance after about 40 epochs. In contrast to our model, the comparison models need more epochs to achieve stable performance. This result shows that our approach can utilize  pre-trained parameters to help the model converge faster and achieve better performance.

\section{Conclusion}
In this paper, we proposed a self-supervised sequential recommendation model S$^3$-Rec based on the mutual information maximization (MIM) principle. In our approach, we adopted the self-attentive recommender architecture as the base model and devised four self-supervised learning objectives to learn the correlations within the raw data.
Based on MIM, the four objectives can learn the correlations among attribute, item, segment, and sequence, which enhances the data representations for sequential recommendation.
Experimental results have shown that our approach outperforms several competitive baselines. 

In the future, we will investigate how to design other forms of self-supervised optimization objectives. 
We will also consider applying our approach 
to more complex recommendation tasks, such as conversational recommendation and multimedia recommendation.

\section*{Acknowledgement}
This work was partially supported by the National Natural Science Foundation of China under Grant No. 61872369 and 61832017,  Beijing Academy of Artificial Intelligence (BAAI) under Grant No. BAAI2020ZJ0301, and Beijing Outstanding Young Scientist Program under Grant No. BJJWZYJH012019100020098, the Fundamental Research Funds for the Central Universities, the Research Funds of Renmin University of China under Grant No.18XNLG22 and 19XNQ047. Xin Zhao is the corresponding author.

\bibliographystyle{ACM-Reference-Format}
\bibliography{sample-base}


\begin{thebibliography}{30}


\ifx \showCODEN    \undefined \def \showCODEN     #1{\unskip}     \fi
\ifx \showDOI      \undefined \def \showDOI       #1{#1}\fi
\ifx \showISBNx    \undefined \def \showISBNx     #1{\unskip}     \fi
\ifx \showISBNxiii \undefined \def \showISBNxiii  #1{\unskip}     \fi
\ifx \showISSN     \undefined \def \showISSN      #1{\unskip}     \fi
\ifx \showLCCN     \undefined \def \showLCCN      #1{\unskip}     \fi
\ifx \shownote     \undefined \def \shownote      #1{#1}          \fi
\ifx \showarticletitle \undefined \def \showarticletitle #1{#1}   \fi
\ifx \showURL      \undefined \def \showURL       {\relax}        \fi
\providecommand\bibfield[2]{#2}
\providecommand\bibinfo[2]{#2}
\providecommand\natexlab[1]{#1}
\providecommand\showeprint[2][]{arXiv:#2}

\bibitem[\protect\citeauthoryear{Devlin, Chang, Lee, and Toutanova}{Devlin
  et~al\mbox{.}}{2019}]%
        {DBLP:conf/naacl/DevlinCLT19}
\bibfield{author}{\bibinfo{person}{J. Devlin}, \bibinfo{person}{M.{-}W. Chang},
  \bibinfo{person}{K. Lee}, {and} \bibinfo{person}{K. Toutanova}.}
  \bibinfo{year}{2019}\natexlab{}.
\newblock \showarticletitle{{BERT:} Pre-training of Deep Bidirectional
  Transformers for Language Understanding}. In
  \bibinfo{booktitle}{\emph{{NAACL-HLT} 2019}}. \bibinfo{pages}{4171--4186}.
\newblock


\bibitem[\protect\citeauthoryear{Gutmann and Hyv{\"{a}}rinen}{Gutmann and
  Hyv{\"{a}}rinen}{2012}]%
        {DBLP:journals/jmlr/GutmannH12}
\bibfield{author}{\bibinfo{person}{M. Gutmann} {and} \bibinfo{person}{A.
  Hyv{\"{a}}rinen}.} \bibinfo{year}{2012}\natexlab{}.
\newblock \showarticletitle{Noise-Contrastive Estimation of Unnormalized
  Statistical Models, with Applications to Natural Image Statistics}.
\newblock \bibinfo{journal}{\emph{J. Mach. Learn. Res.}}  \bibinfo{volume}{13}
  (\bibinfo{year}{2012}), \bibinfo{pages}{307--361}.
\newblock


\bibitem[\protect\citeauthoryear{Hidasi, Karatzoglou, Baltrunas, and
  Tikk}{Hidasi et~al\mbox{.}}{2016a}]%
        {DBLP:journals/corr/HidasiKBT15}
\bibfield{author}{\bibinfo{person}{B. Hidasi}, \bibinfo{person}{A.
  Karatzoglou}, \bibinfo{person}{L. Baltrunas}, {and} \bibinfo{person}{D.
  Tikk}.} \bibinfo{year}{2016}\natexlab{a}.
\newblock \showarticletitle{Session-based Recommendations with Recurrent Neural
  Networks}. In \bibinfo{booktitle}{\emph{{ICLR} 2016}}.
\newblock


\bibitem[\protect\citeauthoryear{Hidasi, Quadrana, Karatzoglou, and
  Tikk}{Hidasi et~al\mbox{.}}{2016b}]%
        {DBLP:conf/recsys/HidasiQKT16}
\bibfield{author}{\bibinfo{person}{B. Hidasi}, \bibinfo{person}{M. Quadrana},
  \bibinfo{person}{A. Karatzoglou}, {and} \bibinfo{person}{D. Tikk}.}
  \bibinfo{year}{2016}\natexlab{b}.
\newblock \showarticletitle{Parallel Recurrent Neural Network Architectures for
  Feature-rich Session-based Recommendations}. In
  \bibinfo{booktitle}{\emph{RecSys 2016}}. \bibinfo{pages}{241--248}.
\newblock


\bibitem[\protect\citeauthoryear{Hjelm, Fedorov, Lavoie{-}Marchildon, Grewal,
  Bachman, Trischler, and Bengio}{Hjelm et~al\mbox{.}}{2019}]%
        {DBLP:conf/iclr/HjelmFLGBTB19}
\bibfield{author}{\bibinfo{person}{R.~D. Hjelm}, \bibinfo{person}{A. Fedorov},
  \bibinfo{person}{S. Lavoie{-}Marchildon}, \bibinfo{person}{K. Grewal},
  \bibinfo{person}{P. Bachman}, \bibinfo{person}{A. Trischler}, {and}
  \bibinfo{person}{Y. Bengio}.} \bibinfo{year}{2019}\natexlab{}.
\newblock \showarticletitle{Learning deep representations by mutual information
  estimation and maximization}. In \bibinfo{booktitle}{\emph{{ICLR} 2019}}.
\newblock


\bibitem[\protect\citeauthoryear{Huang, Ren, Zhao, He, Wen, and Dong}{Huang
  et~al\mbox{.}}{2019}]%
        {DBLP:conf/wsdm/HuangRZHWD19}
\bibfield{author}{\bibinfo{person}{J. Huang}, \bibinfo{person}{Z. Ren},
  \bibinfo{person}{W.~X. Zhao}, \bibinfo{person}{G. He},
  \bibinfo{person}{J.{-}R. Wen}, {and} \bibinfo{person}{D. Dong}.}
  \bibinfo{year}{2019}\natexlab{}.
\newblock \showarticletitle{Taxonomy-Aware Multi-Hop Reasoning Networks for
  Sequential Recommendation}. In \bibinfo{booktitle}{\emph{{WSDM} 2019}}.
  \bibinfo{pages}{573--581}.
\newblock


\bibitem[\protect\citeauthoryear{Huang, Zhao, Dou, Wen, and Chang}{Huang
  et~al\mbox{.}}{2018}]%
        {DBLP:conf/sigir/HuangZDWC18}
\bibfield{author}{\bibinfo{person}{J. Huang}, \bibinfo{person}{W.~X. Zhao},
  \bibinfo{person}{H. Dou}, \bibinfo{person}{J.{-}R. Wen}, {and}
  \bibinfo{person}{E.~Y. Chang}.} \bibinfo{year}{2018}\natexlab{}.
\newblock \showarticletitle{Improving Sequential Recommendation with
  Knowledge-Enhanced Memory Networks}. In \bibinfo{booktitle}{\emph{{SIGIR}
  2018}}. \bibinfo{pages}{505--514}.
\newblock


\bibitem[\protect\citeauthoryear{Kang and McAuley}{Kang and McAuley}{2018}]%
        {DBLP:conf/icdm/KangM18}
\bibfield{author}{\bibinfo{person}{W.{-}C. Kang} {and} \bibinfo{person}{J.~J.
  McAuley}.} \bibinfo{year}{2018}\natexlab{}.
\newblock \showarticletitle{Self-Attentive Sequential Recommendation}. In
  \bibinfo{booktitle}{\emph{{ICDM} 2018}}. \bibinfo{pages}{197--206}.
\newblock


\bibitem[\protect\citeauthoryear{Kingma and Ba}{Kingma and Ba}{2015}]%
        {DBLP:journals/corr/KingmaB14}
\bibfield{author}{\bibinfo{person}{D.~P. Kingma} {and} \bibinfo{person}{J.
  Ba}.} \bibinfo{year}{2015}\natexlab{}.
\newblock \showarticletitle{Adam: {A} Method for Stochastic Optimization}. In
  \bibinfo{booktitle}{\emph{{ICLR} 2015}}.
\newblock


\bibitem[\protect\citeauthoryear{Kong, de~Masson~d'Autume, Yu, Ling, Dai, and
  Yogatama}{Kong et~al\mbox{.}}{2020}]%
        {DBLP:conf/iclr/KongdYLDY20}
\bibfield{author}{\bibinfo{person}{L. Kong}, \bibinfo{person}{C. de
  Masson~d'Autume}, \bibinfo{person}{L. Yu}, \bibinfo{person}{W. Ling},
  \bibinfo{person}{Z. Dai}, {and} \bibinfo{person}{D. Yogatama}.}
  \bibinfo{year}{2020}\natexlab{}.
\newblock \showarticletitle{A Mutual Information Maximization Perspective of
  Language Representation Learning}. In \bibinfo{booktitle}{\emph{{ICLR}
  2020}}.
\newblock


\bibitem[\protect\citeauthoryear{Linsker}{Linsker}{1988}]%
        {DBLP:journals/computer/Linsker88}
\bibfield{author}{\bibinfo{person}{R. Linsker}.}
  \bibinfo{year}{1988}\natexlab{}.
\newblock \showarticletitle{Self-Organization in a Perceptual Network}.
\newblock \bibinfo{journal}{\emph{{IEEE} Computer}} \bibinfo{volume}{21},
  \bibinfo{number}{3} (\bibinfo{year}{1988}), \bibinfo{pages}{105--117}.
\newblock


\bibitem[\protect\citeauthoryear{Logeswaran and Lee}{Logeswaran and
  Lee}{2018}]%
        {DBLP:conf/iclr/LogeswaranL18}
\bibfield{author}{\bibinfo{person}{L. Logeswaran} {and} \bibinfo{person}{H.
  Lee}.} \bibinfo{year}{2018}\natexlab{}.
\newblock \showarticletitle{An efficient framework for learning sentence
  representations}. In \bibinfo{booktitle}{\emph{{ICLR} 2018}}.
\newblock


\bibitem[\protect\citeauthoryear{Ma, Kang, and Liu}{Ma et~al\mbox{.}}{2019}]%
        {DBLP:conf/kdd/MaKL19}
\bibfield{author}{\bibinfo{person}{C. Ma}, \bibinfo{person}{P. Kang}, {and}
  \bibinfo{person}{X. Liu}.} \bibinfo{year}{2019}\natexlab{}.
\newblock \showarticletitle{Hierarchical Gating Networks for Sequential
  Recommendation}. In \bibinfo{booktitle}{\emph{{KDD} 2019}}.
  \bibinfo{pages}{825--833}.
\newblock


\bibitem[\protect\citeauthoryear{McAuley, Targett, Shi, and van~den
  Hengel}{McAuley et~al\mbox{.}}{2015}]%
        {DBLP:conf/sigir/McAuleyTSH15}
\bibfield{author}{\bibinfo{person}{J.~J. McAuley}, \bibinfo{person}{C.
  Targett}, \bibinfo{person}{Q. Shi}, {and} \bibinfo{person}{A. van~den
  Hengel}.} \bibinfo{year}{2015}\natexlab{}.
\newblock \showarticletitle{Image-Based Recommendations on Styles and
  Substitutes}. In \bibinfo{booktitle}{\emph{{SIGIR} 2015}}.
  \bibinfo{pages}{43--52}.
\newblock


\bibitem[\protect\citeauthoryear{Mikolov, Sutskever, Chen, Corrado, and
  Dean}{Mikolov et~al\mbox{.}}{2013}]%
        {DBLP:conf/nips/MikolovSCCD13}
\bibfield{author}{\bibinfo{person}{T. Mikolov}, \bibinfo{person}{I. Sutskever},
  \bibinfo{person}{K. Chen}, \bibinfo{person}{G.~S. Corrado}, {and}
  \bibinfo{person}{J. Dean}.} \bibinfo{year}{2013}\natexlab{}.
\newblock \showarticletitle{Distributed Representations of Words and Phrases
  and their Compositionality}. In \bibinfo{booktitle}{\emph{{NeurIPS} 2013}}.
  \bibinfo{pages}{3111--3119}.
\newblock


\bibitem[\protect\citeauthoryear{Pasricha and McAuley}{Pasricha and
  McAuley}{2018}]%
        {DBLP:conf/recsys/PasrichaM18}
\bibfield{author}{\bibinfo{person}{R. Pasricha} {and} \bibinfo{person}{J.~J.
  McAuley}.} \bibinfo{year}{2018}\natexlab{}.
\newblock \showarticletitle{Translation-based factorization machines for
  sequential recommendation}. In \bibinfo{booktitle}{\emph{{RecSys} 2018}}.
  \bibinfo{pages}{63--71}.
\newblock


\bibitem[\protect\citeauthoryear{Quadrana, Karatzoglou, Hidasi, and
  Cremonesi}{Quadrana et~al\mbox{.}}{2017}]%
        {DBLP:conf/recsys/QuadranaKHC17}
\bibfield{author}{\bibinfo{person}{M. Quadrana}, \bibinfo{person}{A.
  Karatzoglou}, \bibinfo{person}{B. Hidasi}, {and} \bibinfo{person}{P.
  Cremonesi}.} \bibinfo{year}{2017}\natexlab{}.
\newblock \showarticletitle{Personalizing Session-based Recommendations with
  Hierarchical Recurrent Neural Networks}. In
  \bibinfo{booktitle}{\emph{{RecSys} 2017}}. \bibinfo{pages}{130--137}.
\newblock


\bibitem[\protect\citeauthoryear{Ren, Chen, Li, Ren, Ma, and de~Rijke}{Ren
  et~al\mbox{.}}{2019}]%
        {DBLP:conf/aaai/RenCLR0R19}
\bibfield{author}{\bibinfo{person}{Pengjie Ren}, \bibinfo{person}{Zhumin Chen},
  \bibinfo{person}{Jing Li}, \bibinfo{person}{Zhaochun Ren},
  \bibinfo{person}{Jun Ma}, {and} \bibinfo{person}{Maarten de Rijke}.}
  \bibinfo{year}{2019}\natexlab{}.
\newblock \showarticletitle{RepeatNet: {A} Repeat Aware Neural Recommendation
  Machine for Session-Based Recommendation}. In
  \bibinfo{booktitle}{\emph{{AAAI} 2019}}. \bibinfo{pages}{4806--4813}.
\newblock


\bibitem[\protect\citeauthoryear{Ren, Liu, Li, Zhao, Wang, Ding, and Wen}{Ren
  et~al\mbox{.}}{2020}]%
        {DBLP:conf/sigir/RenLLZWDW20}
\bibfield{author}{\bibinfo{person}{R. Ren}, \bibinfo{person}{Z. Liu},
  \bibinfo{person}{Y. Li}, \bibinfo{person}{W.~X. Zhao}, \bibinfo{person}{H.
  Wang}, \bibinfo{person}{B. Ding}, {and} \bibinfo{person}{J.{-}R. Wen}.}
  \bibinfo{year}{2020}\natexlab{}.
\newblock \showarticletitle{Sequential Recommendation with Self-Attentive
  Multi-Adversarial Network}. In \bibinfo{booktitle}{\emph{{SIGIR} 2020}}.
  \bibinfo{pages}{89--98}.
\newblock


\bibitem[\protect\citeauthoryear{Rendle}{Rendle}{2010}]%
        {DBLP:conf/icdm/Rendle10}
\bibfield{author}{\bibinfo{person}{S. Rendle}.}
  \bibinfo{year}{2010}\natexlab{}.
\newblock \showarticletitle{Factorization Machines}. In
  \bibinfo{booktitle}{\emph{{ICDM} 2010}}. \bibinfo{pages}{995--1000}.
\newblock


\bibitem[\protect\citeauthoryear{Rendle, Freudenthaler, and
  Schmidt{-}Thieme}{Rendle et~al\mbox{.}}{2010}]%
        {DBLP:conf/www/RendleFS10}
\bibfield{author}{\bibinfo{person}{S. Rendle}, \bibinfo{person}{C.
  Freudenthaler}, {and} \bibinfo{person}{L. Schmidt{-}Thieme}.}
  \bibinfo{year}{2010}\natexlab{}.
\newblock \showarticletitle{Factorizing personalized Markov chains for
  next-basket recommendation}. In \bibinfo{booktitle}{\emph{{WWW} 2010}}.
  \bibinfo{pages}{811--820}.
\newblock


\bibitem[\protect\citeauthoryear{Song, Shi, Xiao, Duan, Xu, Zhang, and
  Tang}{Song et~al\mbox{.}}{2019}]%
        {DBLP:conf/cikm/SongS0DX0T19}
\bibfield{author}{\bibinfo{person}{W. Song}, \bibinfo{person}{C. Shi},
  \bibinfo{person}{Z. Xiao}, \bibinfo{person}{Z. Duan}, \bibinfo{person}{Y.
  Xu}, \bibinfo{person}{M. Zhang}, {and} \bibinfo{person}{J. Tang}.}
  \bibinfo{year}{2019}\natexlab{}.
\newblock \showarticletitle{AutoInt: Automatic Feature Interaction Learning via
  Self-Attentive Neural Networks}. In \bibinfo{booktitle}{\emph{{CIKM} 2019}}.
  \bibinfo{pages}{1161--1170}.
\newblock


\bibitem[\protect\citeauthoryear{Sun, Liu, Wu, Pei, Lin, Ou, and Jiang}{Sun
  et~al\mbox{.}}{2019}]%
        {DBLP:conf/cikm/SunLWPLOJ19}
\bibfield{author}{\bibinfo{person}{F. Sun}, \bibinfo{person}{J. Liu},
  \bibinfo{person}{J. Wu}, \bibinfo{person}{C. Pei}, \bibinfo{person}{X. Lin},
  \bibinfo{person}{W. Ou}, {and} \bibinfo{person}{P. Jiang}.}
  \bibinfo{year}{2019}\natexlab{}.
\newblock \showarticletitle{BERT4Rec: Sequential Recommendation with
  Bidirectional Encoder Representations from Transformer}. In
  \bibinfo{booktitle}{\emph{{CIKM} 2019}}. \bibinfo{pages}{1441--1450}.
\newblock


\bibitem[\protect\citeauthoryear{Tang and Wang}{Tang and Wang}{2018}]%
        {DBLP:conf/wsdm/TangW18}
\bibfield{author}{\bibinfo{person}{J. Tang} {and} \bibinfo{person}{K. Wang}.}
  \bibinfo{year}{2018}\natexlab{}.
\newblock \showarticletitle{Personalized Top-N Sequential Recommendation via
  Convolutional Sequence Embedding}. In \bibinfo{booktitle}{\emph{{WSDM}
  2018}}. \bibinfo{pages}{565--573}.
\newblock


\bibitem[\protect\citeauthoryear{van~den Oord, Li, and Vinyals}{van~den Oord
  et~al\mbox{.}}{2018}]%
        {DBLP:journals/corr/abs-1807-03748}
\bibfield{author}{\bibinfo{person}{A. van~den Oord}, \bibinfo{person}{Y. Li},
  {and} \bibinfo{person}{O. Vinyals}.} \bibinfo{year}{2018}\natexlab{}.
\newblock \showarticletitle{Representation Learning with Contrastive Predictive
  Coding}.
\newblock \bibinfo{journal}{\emph{CoRR}}  \bibinfo{volume}{abs/1807.03748}
  (\bibinfo{year}{2018}).
\newblock
\showeprint[arxiv]{1807.03748}


\bibitem[\protect\citeauthoryear{Vaswani, Shazeer, Parmar, Uszkoreit, Jones,
  Gomez, Kaiser, and Polosukhin}{Vaswani et~al\mbox{.}}{2017}]%
        {DBLP:conf/nips/VaswaniSPUJGKP17}
\bibfield{author}{\bibinfo{person}{A. Vaswani}, \bibinfo{person}{N. Shazeer},
  \bibinfo{person}{N. Parmar}, \bibinfo{person}{J. Uszkoreit},
  \bibinfo{person}{L. Jones}, \bibinfo{person}{A.~N. Gomez},
  \bibinfo{person}{L. Kaiser}, {and} \bibinfo{person}{I. Polosukhin}.}
  \bibinfo{year}{2017}\natexlab{}.
\newblock \showarticletitle{Attention is All you Need}. In
  \bibinfo{booktitle}{\emph{{NeurIPS} 2017}}. \bibinfo{pages}{5998--6008}.
\newblock


\bibitem[\protect\citeauthoryear{Xin, Karatzoglou, Arapakis, and Jose}{Xin
  et~al\mbox{.}}{2020}]%
        {DBLP:conf/sigir/XinKAJ20}
\bibfield{author}{\bibinfo{person}{Xin Xin}, \bibinfo{person}{Alexandros
  Karatzoglou}, \bibinfo{person}{Ioannis Arapakis}, {and}
  \bibinfo{person}{Joemon~M. Jose}.} \bibinfo{year}{2020}\natexlab{}.
\newblock \showarticletitle{Self-Supervised Reinforcement Learning for
  Recommender Systems}. In \bibinfo{booktitle}{\emph{{SIGIR} 2020}}.
  \bibinfo{pages}{931--940}.
\newblock


\bibitem[\protect\citeauthoryear{Yeh and Chen}{Yeh and Chen}{2019}]%
        {DBLP:conf/emnlp/YehC19}
\bibfield{author}{\bibinfo{person}{Y.{-}T. Yeh} {and} \bibinfo{person}{Y.{-}N.
  Chen}.} \bibinfo{year}{2019}\natexlab{}.
\newblock \showarticletitle{QAInfomax: Learning Robust Question Answering
  System by Mutual Information Maximization}. In
  \bibinfo{booktitle}{\emph{{EMNLP-IJCNLP} 2019}}. \bibinfo{pages}{3368--3373}.
\newblock


\bibitem[\protect\citeauthoryear{Zhang, Zhao, Liu, Sheng, Xu, Wang, Liu, and
  Zhou}{Zhang et~al\mbox{.}}{2019}]%
        {DBLP:conf/ijcai/ZhangZLSXWLZ19}
\bibfield{author}{\bibinfo{person}{T. Zhang}, \bibinfo{person}{P. Zhao},
  \bibinfo{person}{Y. Liu}, \bibinfo{person}{V.~S. Sheng}, \bibinfo{person}{J.
  Xu}, \bibinfo{person}{D. Wang}, \bibinfo{person}{G. Liu}, {and}
  \bibinfo{person}{X. Zhou}.} \bibinfo{year}{2019}\natexlab{}.
\newblock \showarticletitle{Feature-level Deeper Self-Attention Network for
  Sequential Recommendation}. In \bibinfo{booktitle}{\emph{{IJCAI} 2019}}.
  \bibinfo{pages}{4320--4326}.
\newblock


\bibitem[\protect\citeauthoryear{Zhou, Zhao, Bian, Zhou, Wen, and Yu}{Zhou
  et~al\mbox{.}}{2020}]%
        {DBLP:journals/corr/abs-2007-04032}
\bibfield{author}{\bibinfo{person}{Kun Zhou}, \bibinfo{person}{Wayne~Xin Zhao},
  \bibinfo{person}{Shuqing Bian}, \bibinfo{person}{Yuanhang Zhou},
  \bibinfo{person}{Ji{-}Rong Wen}, {and} \bibinfo{person}{Jingsong Yu}.}
  \bibinfo{year}{2020}\natexlab{}.
\newblock \showarticletitle{Improving Conversational Recommender Systems via
  Knowledge Graph based Semantic Fusion}. In \bibinfo{booktitle}{\emph{KDD
  2020}}.
\newblock


\end{thebibliography}

\appendix


\end{document}